\documentclass[
reprint,
superscriptaddress,
frontmatterverbose, 
showpacs,
preprintnumbers,
nofootinbib,
amsmath,
amssymb,
aps,
floatfix,
twocolumn
]{revtex4-2}

\usepackage[normalem]{ulem}
\usepackage{graphicx}
\usepackage{dcolumn}
\usepackage{array} 
\usepackage{bm}
\usepackage{color}
\usepackage[dvipsnames]{xcolor}
\usepackage[
    colorlinks = true,
    linkcolor = BlueViolet,
    anchorcolor = purple,
    citecolor = magenta,
    filecolor = purple,
    urlcolor = magenta]{hyperref}
\usepackage{graphicx}    
\usepackage{url}
\usepackage{xspace}
\usepackage{slashed}
\usepackage{multirow,bigstrut}
\usepackage{mathrsfs} 
\usepackage{relsize,amsmath}
\usepackage[geometry]{ifsym}
\usepackage{amssymb}
\usepackage{pifont}
\usepackage{physics}
\usepackage[compat=1.1.0]{tikz-feynman}
\usepackage{cleveref}

\usepackage{rotating}
\usepackage{ragged2e}
\usepackage{slashed}
\usepackage{longtable}

\usepackage{fontawesome} 
\definecolor{blue-violet}{rgb}{0.33, 0.17, 0.89}

\renewcommand{\phi}{\varphi}

\newcounter{CommentCount}
\definecolor{MH}{rgb}{0.0,0.6,9}
\definecolor{palatinate}{rgb}{0.494, 0.192, 0.482}

\definecolor{teal}{HTML}{008080}

\renewcommand{\arraystretch}{1.3}
\usepackage{siunitx}

\DeclareSIUnit \s {\second}
\DeclareSIUnit \ns {\nano\second}
\DeclareSIUnit \mus {\micro\second}
\DeclareSIUnit \ms {\milli\second}
\DeclareSIUnit \MB {\mega\byte}
\DeclareSIUnit \GB {\giga\byte}
\DeclareSIUnit \TB {\tera\byte}
\DeclareSIUnit \PB {\peta\byte}
\DeclareSIUnit \Mbps {\mega\bit/\s}
\DeclareSIUnit \Gbps {\giga\bit/\s}
\DeclareSIUnit \Tbps {\tera\bit/\s}
\DeclareSIUnit \Pbps {\peta\bit/\s}
\DeclareSIUnit \kton {\kilo\tonne} 
\DeclareSIUnit \kt {\kilo\tonne}
\DeclareSIUnit \Mt {\mega\tonne}
\DeclareSIUnit \eV {\electronvolt}
\DeclareSIUnit \keV {\kilo\electronvolt}
\DeclareSIUnit \MeV {\mega\electronvolt}
\DeclareSIUnit \GeV {\giga\electronvolt}
\DeclareSIUnit \TeV {\tera\electronvolt}
\DeclareSIUnit \PeV {\peta\electronvolt}
\DeclareSIUnit \EeV {\exa\electronvolt}
\DeclareSIUnit \m {\meter}
\DeclareSIUnit \cm {\centi\meter}
\DeclareSIUnit \in {\inchcommand}
\DeclareSIUnit \km {\kilo\meter}
\DeclareSIUnit \kV {\kilo\volt}
\DeclareSIUnit \kW {\kilo\watt}
\DeclareSIUnit \MW {\mega\watt}
\DeclareSIUnit \MHz {\mega\hertz}
\DeclareSIUnit \mrad {\milli\radian}
\DeclareSIUnit \year {years}
\DeclareSIUnit \POT {POT}
\DeclareSIUnit \sig {$\sigma$}
\DeclareSIUnit\parsec{pc}
\DeclareSIUnit\lightyear{ly}
\DeclareSIUnit\foot{ft}
\DeclareSIUnit\ft{ft}
\DeclareSIUnit \ppb{ppb}
\DeclareSIUnit \ppt{ppt}
\DeclareSIUnit \samples{S}
\DeclareSIUnit \pe{PE}
\DeclareSIUnit \T{T}

\newcommand{\enu}{\E_\enu}

\newcommand\response[1]{{\color{black}  #1}}

\definecolor{myred}{cmyk}{0,1,1,0.55}
\definecolor{mygreen}{rgb}{0.27, 0.64, 0.48}
\definecolor{mygray}{gray}{.95}

\begin{document}
\hfill {\tt MITP-25-011}
\title{\Large  Single pion resonant production in BSM scenarios: cross sections and amplitudes} 
\author{Jaime Hoefken Zink}
\affiliation{National Centre for Nuclear Research, Pasteura 7, Warsaw, PL-02-093, Poland}
\email{jaime.hoefkenzink@ncbj.gov.pl}


\author{Maura E. Ramirez-Quezada}
\affiliation{PRISMA$^+$ Cluster of Excellence \& Mainz Institute for Theoretical Physics,\\ Johannes Gutenberg University, 55099 Mainz, Germany}
\email{mramirez@uni-mainz.de}
\affiliation{Dual CP Institute of High Energy Physics, C.P. 28045, Colima, M\'exico.}

\begin{abstract}
We present a comprehensive theoretical framework describing single pion resonant production through inelastic dark fermion–nucleon interactions mediated by resonances in the GeV-scale regime. Building upon the Rein–Sehgal approach, we derive differential cross sections for processes in which an incoming dark fermion scatters off a nucleon, exciting a resonance that subsequently decays into a nucleon and a pion. Our formulation accommodates various mediator types—namely, dark photons with vector and axial couplings, as well as scalar and pseudoscalar mediators—thereby extending the conventional approach that Rein, Sehgal and Berger performed for neutrino interactions. Transition amplitudes for the nucleon-to-resonance conversion are computed using the relativistic harmonic-oscillator quark model from Feynman, Kislinger and Ravndal, while a Breit–Wigner prescription is employed to incorporate finite resonance widths. This framework offers a useful tool for interpreting experimental data in dark sector and dark matter searches and represents a contribution to elucidate the possible role of resonances in GeV-scale phenomenology.
\end{abstract}
\preprint{}
\maketitle

\section{Introduction}

Resonances at GeV-scale energies bridge the gap between quasi-elastic processes and deep inelastic scattering. Studying resonant cross sections is essential for probing new physics -- discrepancies between predicted and observed cross sections may reveal novel interactions or the presence of exotic particles, such as dark matter (DM) candidates, thereby opening a promising window into phenomena that extend our current understanding of particle physics. Despite their clear experimental signatures and potential relevance for beyond the Standard Model (BSM) physics, this resonant regime remains comparatively unexplored \response{probably due to the lack of BSM experimental data to model and fit the interactions}. 

The richness of recent studies on boosted DM is a strong motivation for performing better searches on how dark sector particles could interact with ordinary matter in the inelastic and even the deeply inelastic regime~\cite{Bringmann:2018cvk, Wang:2021jic, Granelli:2022ysi, COSINE-100:2023tcq, Aoki:2023tlb, Cappiello:2024acu, Herbermann:2024kcy, Kim:2024ltz, HoefkenZink:2024hor, Nagao:2024hit, Ghosh:2024dqw, ICARUS:2024lew, Lu:2024xxb, PandaX:2024syk, DeMarchi:2024riu, DeMarchi:2024zer, Sun:2025gyj, Jeesun:2025gzt, BetancourtKamenetskaia:2025noa, Liang:2025etf}. The energies at which DM particles can be accelerated~\cite{Bringmann:2018cvk, Wang:2021jic} can be very sensitive to resonant inelastic scattering. Furthermore, dark sectors, sub-GeV extensions of the SM that interact with it through the so called ``portals"~\cite{Okawa:2019arp, Ballett:2019pyw, Darme:2020ral, Wojcik:2020wgm, Contino:2020tix, Demir:2021yii, Abdullahi:2020nyr, Aboubrahim:2022bzk, Costa:2022pxv, Biondini:2023yxt, Carrasco:2023loy, Abdullahi:2023tyk, Abdullahi:2023ejc, CMS:2024zqs}, can also be present in experiments in interactions where the energies are relevant for resonant processes. The particular signature of a pion coming out from a hadronic vertex could be used to explore BSM signatures beyond what has been done so far. Therefore, an estimation of the interaction rates in this resonant region is of considerable importance.

Single-pion production in the GeV-scale resonance region has been extensively studied since Adler's unified approach to photon-, electro-, and weak processes~\cite{ADLER1968189}. In his work, Adler extended the framework of pion photoproduction to include both electro- and weak production. His pioneering approach established key kinematic relations, addressed gauge invariance issues, and incorporated vector and axial-vector contributions, thereby laying a foundational framework for subsequent research, as evidenced by numerous partial-wave analyses of pion photoproduction~\cite{Moorhouse:1973mr,PhysRevD.9.1,Metcalf:1974ij,Knies:1974wg}.

In order to compute the transition amplitude between the nucleons and the resonances, a pioneer approach is the relativistic harmonic-oscillator quark model of Feynman, Kislinger and Ravndal (FKR)~\cite{Feynman:1971wr,Ravndal:1971cuf,Ravndal:1973xx}, which interprets hadrons as a 4D-harmonic oscillator excitations and provides a framework for calculating transitions among hadrons in the Standard Model (SM). Building on FKR, Rein and Sehgal adapted the model to neutrino-induced single-pion production, introducing an adjustable axial mass parameter and practical algorithms for Monte Carlo event generation~\cite{Rein:1980wg}. Rein further examined angular distribution in neutrino-induced single-pion channels~\cite{Rein:1987}. Originally, the model considered vanishing masses on the leptonic side. Finite lepton masses were included into the Rein-Sehgal model (RS) to address discrepancies for single pion neutrino-production by \cite{Kuzmin:2003ji} and deficits arising from lepton mass corrections due to the pion-pole~\cite{Berger:2007rq, Graczyk:2007xk}. \response{This approach, while not being the most updated one, allows one to predict similar interactions but with BSM new fields with different mediators.}
%

In this work, we present a comprehensive and unified framework for calculating dark sectors–nucleon inelastic interactions that produce a single pion in the final state, through the nearly on-shell production of a $N-$ or $\Delta-$resonance. Our approach derives differential cross sections by establishing the kinematics in both the Lab and isobaric (IB) frames and employs the Rein–Sehgal method to model the non-hadronic current~\cite{Rein:1980wg}. Among the new features with respect to previous computations, we include the effects of the masses of all the particles involved, we regard a mediator whose mass is not necessarily much heavier than the transferred momentum, such as it is the case with the SM weak interactions, and we include the interaction not only through vector-axial mediators, but also through scalar and pseudoscalar ones, leaving all the couplings free. These improvements enable a consistent calculation of cross sections and amplitudes. Our results shed lights on the role of resonances in the GeV-scale phenomenology and open new avenues for experimental searches. 

Although  the FKR and Rein-Sehgal frameworks are not the most modern methods for computing resonant cross sections---and experimental corrections are necessary for high precision---their simplicity and adaptability to BSM models make them invaluable for initial studies. This is specially true in  astrophysical scenarios, where a reliable  estimation of cross sections is often sufficient compared to the level of precision demanded by terrestrial detectors. 
%

The paper is organized as follows, in \Cref{sec:cross-section}, we present the theoretical formulation to calculate the cross sections for inelastic interactions between dark fermions and nucleons via resonances. \Cref{sec:form-factors} outlines the calculation of nucleon-resonance transition amplitudes using the FKR quark model. A set of explicit expressions for vector, axial, scalar and pseudoscalar amplitudes essential for modeling resonant processes are given. We summarized in \Cref{sec:summary}.

\section{Cross sections}
\label{sec:cross-section}

\begin{figure}
\begin{tikzpicture}[baseline=(current bounding box.center)]
\begin{feynman} 
\vertex(V1);
\vertex(p1)[above left=1.5cm of V1];
\vertex(p3)[above right=1.5cm of V1];
\vertex[below = 1.5cm of V1](V2);
\vertex(p4)[right=2cm of V2];
\vertex(p2)[below left=1.5cm of V2];
\vertex[right=2cm of V2](V3);
\vertex(p5)[above right=1.5cm of V3];
\vertex(p6)[below right=1.5cm of V3];
\diagram*{
(p1)--[fermion, edge label'=$\chi(p_1)$](V1),
(p2)--[fermion, edge label'=$N(p_2)$](V2),
(V1)--[boson, edge label'=$Z^\prime / \Phi / a (q)$](V2),
(V1) --[fermion, edge label'=$\chi(p_3)$](p3),
(V2) --[fermion, edge label=$N^*(p_4)$](p4),
(V3) --[fermion, edge label'=$N'(k_1)$](p5),
(V3) --[fermion, edge label'=$\pi(k_2)$](p6)
}; 
\end{feynman}
\end{tikzpicture}
\caption{\justifying DM-nucleon inelastic interaction with a dark photon ($Z'$)/scalar ($\Phi$) / pseudoscalar ($a$) mediator that produces a $N-$ or $\Delta-$ resonance ($N^*$) that further decays into a nucleon and a pion.}\label{diagram:DM_RES}
\end{figure}
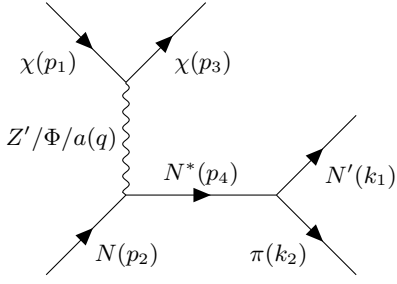

In this section, we derive the cross sections for inelastic interactions between dark fermions and nucleons that are mediated by intermediate resonances.
In order to quantify these interactions via resonances, we consider the process
\begin{equation}
    \chi(p_1) N(p_2) \to \chi(p_3) N^*(p_4) \to \chi(p_3) N'(k_1) \pi (k_2)\,,
\end{equation}
 where $\chi$ is the dark fermion, $N$ or $N'$ are nucleons and $N^*$ is a GeV-scale $N-$ or $\Delta-$resonance, as Illustrated in \Cref{diagram:DM_RES}. \response{This resonance must have the same charge of the target nucleon.}  Here, $N(p_2)$ is assumed to be at rest in the Lab frame. The interaction is mediated by either a dark photon (with both, vector and axial couplings), a scalar, or a pseudoscalar, all of which are electromagnetically neutral. Consequently, four distinct channels arise, 
\begin{enumerate}
    \item $\chi + p \to \chi + p + \pi^0$,
    \item $\chi + p \to \chi + n + \pi^+$,
    \item $\chi + n \to \chi + n + \pi^0$,
    \item $\chi + n \to \chi + p + \pi^-$.
\end{enumerate}
We adopt a model-independent approach to new physics, outlining only the 
interaction forms for each mediator. These interactions are parametrized as follows: For the vector-axial case the Lagrangian is,
\begin{equation}
\label{eq:Lagr_v-a}
\begin{split}
\mathcal{L}_I^\mathrm{V-A} = &\sum_{f \in SM} g_{f Z^\prime} \, \overline{f} \, \gamma^\mu \left( c_V^f - c_A^f \, \gamma^5 \right) f \, Z^\prime_\mu \\
&+ g_{D} \, \overline{\chi} \, \gamma^\mu \left( c_V^\chi - c_A^\chi \, \gamma^5 \right) \chi \, Z^\prime_\mu\,,
\end{split}
\end{equation}
while for the scalar we have,
\begin{equation}
\label{eq:Lagr_S}
\begin{split}
\mathcal{L}_I^\mathrm{S} = &\sum_{f \in SM} g_{f \Phi} \, \overline{f} \, f \, \Phi \;+\; g_{D} \, \overline{\chi} \, \chi \, \Phi\,.
\end{split}
\end{equation}
and the pseudoscalar,
\begin{equation}
\label{eq:Lagr_P}
\begin{split}
\mathcal{L}_I^\mathrm{P} = &-i\sum_{f \in SM} g_{f a} \overline{f} \gamma^5 f a - i g_{D} \overline{\chi} \gamma^5 \chi a\,,
\end{split}
\end{equation}
such that $Z^\prime$ is the boson with vector-axial ($V-A$) couplings, $\Phi$ is a scalar and $a$, a pseudoscalar. $f \in SM$ means that $f$ is a fermion of the Standard Model of particle physics. We are also considering one kind of dark sector fermions, $\chi$, which have a coupling $g_D$ to the dark mediators. Since we will just be considering quarks in the SM, we will use a particular parametrization of their couplings, so that we can have an overall factor that accounts for the coupling order with the quarks in the nucleons, $g_{NM}$- being $M$ the mediator, and a coefficient that turns the previous factor into the coupling itself for each quark, $g_q^k$ - being $q$ the quark and $k$ the kind of coupling (V, A, S, P). The relations between the couplings in \Cref{eq:Lagr_v-a,eq:Lagr_S,eq:Lagr_P} and the ones that will be used in the rest of the paper are:
\begin{equation}
\begin{split}
g_{q Z^\prime} \times c_V^q &\to g_{N Z^\prime} \times g_V^q\,,\\
g_{q \Phi} &\to g_{N \Phi} \times g_S^q\,,\\
g_{q a} &\to g_{N a} \times g_P^q\,.
\end{split}
\end{equation}

We follow the approach of Rein and Sehgal~\cite{Rein:1980wg}, by treating the non-hadronic current matrix element as the polarization vector of the virtual intermediate boson. In vector-mediated processes, this element is further contracted with the  propagator's vector structure. For scalar and pseudoscalar interactions, although the matrix element is not a vector, it still serves as the mediator. We refer to this as the ``matrix element mediator" (MEM).
%

\subsection{Kinematics in the resonant frame}
The formalism introduced by Ravndal~\cite{Ravndal:1973xx} and later refined by Rein and Sehgal~\cite{Rein:1980wg} is constructed in  the resonant, IB frame. In this frame, the resonance remains at rest, the  momentum transfer $q$ is aligned with the  $\hat{z}$ axis, the $\hat{y}$ axis is defined along  the direction of $\vec{p}_1 \times \vec{p}_3$, and  the $\hat{x}$ axis follow $\left( \vec{p}_1 \times \vec{p}_3 \right) \times \vec{q}$, where $q \equiv p_1 - p_3$.
In the Lab frame, where the target nucleon $N(p_2)$ is at rest, the kinematical variables are given by
\begin{equation}
\begin{split}
&p_1 = \left( E_1, 0, 0, |\vec{p}_1|\right),\\[0.05cm]
&p_2 = \left( m_N, 0, 0, 0\right),\\[0.05cm]
&p_3 = \left( E_3, |\vec{p}_3| \sin \theta, 0, |\vec{p}_3| \cos \theta\right),\\[0.05cm]
&p_4 = p_1 + p_2 - p_3,\\[0.05cm]
&q = p_1 - p_2\,.
\end{split}
\end{equation}
\begin{widetext}
Moving to the IB frame,\footnote{If not explicitly stated, all parameters are in the Lab frame} we obtain:
\begin{equation}
\begin{split}
&p_1^\mathrm{IB} = \left(\frac{(E_1 - E_3)^2 + 2 m_N E_1 - Q^2}{2W} , A_{13}, 0, B_{13}^-\right),\\[0.1cm]
&p_2^\mathrm{IB} = \left(\frac{m_N(E_1 - E_3 + m_N)}{W}, 0, 0, -\frac{ m_N Q }{W}\right),\\[0.1cm]
&p_3^\mathrm{IB} = \left(-\frac{(E_1 - E_3)^2 - 2 m_N E_3 - Q^2}{2W}, A_{13}, 0, B_{13}^+\right),\\[0.1cm]
&p_4^\mathrm{IB} = \left(W, 0, 0, 0\right),\\[0.1cm]
&A_{13} =  \frac{\sqrt{\left(|\vec{p}_1| + |\vec{p}_3| - Q\right)\left(|\vec{p}_1| - |\vec{p}_3| + Q\right)\left(-|\vec{p}_1| + |\vec{p}_3| + Q\right)\left(|\vec{p}_1| + |\vec{p}_3| + Q\right))} }{2Q}, \\[0.1cm]
&B_{13}^{\pm} = \frac{(E_1^2 - E_3^2)(E_1 - E_3 + m_N) - (E_1 + E_3 \pm m_N) Q^2}{2QW}\,,
\end{split}
\end{equation}
\end{widetext}
where $W$ is the invariant mass of the outgoing nucleon and pion. In the literature, the momentum transfer in the IB fame , is usually denoted as  $q^\mathrm{IB}=\left( \nu^*, 0, 0, Q^* \right)$, with 
\begin{equation}
    Q^* =\frac{m_N |\vec{q}|}{W}\,
\end{equation}
 and 
 \begin{equation}
\nu^* =     \frac{(E_1 - E_3)(E_1- E_3 + m_N) - |\vec{q}|^2}{W}.
 \end{equation}

\subsection{Differential cross section for an interaction through a dark photon}
In the case of a dark photon mediator, the amplitude for the process shown in \Cref{diagram:DM_RES}---in which  an incoming dark fermion $\chi(p_1,\lambda_1)$ and nucleon $N(p_2)$ interact to produce an outgoing dark fermion  $\chi(p_3,\lambda_2)$ and nucleon resonance $N^*(p_4)$---is given by, 
\begin{align}
&\mathcal{M} (\chi(p_1, \lambda_1) N(p_2) \to \chi(p_3, \lambda_2) N^{*} (p_4))\nonumber\\[0.1cm] &= \frac{g_D g_{NZ'}}{q^2 - m^2_{Z^\prime}} \left[ \overline{u}_{p_3 \lambda_2} \gamma_\mu \left( c_V^\chi - c_A^\chi \gamma^5 \right) u_{p_1 \lambda_1} \right]\times\nonumber \\
&  \times \left( g^{\mu \nu} - q^{\mathrm{IB} \mu} q^{\mathrm{IB} \nu} / m^2_{Z^\prime}  \right)  \bra{N^{*}} J_\nu^{V+} (0) - J_\nu^{A+} (0) \ket{N}\nonumber \\[0.2cm]
&= \!\!2M\! \frac{g_D g_{ NZ'}}{q^2 - m^2_{Z^\prime}} \!V_{\lambda_1 \lambda_2}^\mu \!( \bra{N^{*}} \!F_\mu^{V}\!\! \ket{N} \!-\! \bra{N^{*}}\! F_\mu^{A}\!\ket{N})
\end{align}
where $F_\mu^{V(A)} \equiv J_\nu^{V(A)+} (0) / (2M)$ are the vector (axial) amplitudes for the nucleon-to-resonance transition, and  $M$ is the resonance mass. In our approach, we set  $M$ equal to the invariant mass $W$ of the outgoing nucleon-pion system. This identification allows us to replace the on-shell narrow decay width with a finite width description by introducing Breit-Winger factor, since $W$ is not exactly equal to $M$. The non-hadronic current, or MEM, for the vector-axial interaction is defined as,
\begin{equation}
  \!\!  V_{\lambda_1 \lambda_2}^{\nu }\!\!\equiv\!\! \left[\bar{u}_{\chi ,\lambda _2} \!\gamma _{\mu } \! \left(c_V^{\chi }-\gamma ^5 c_A^{\chi }\right) u_{\chi ,\lambda _1} \right] \!\!\!\left(\!g^{\mu \nu }\!\!-\!\frac{q^{\mathrm{IB} \mu} q^{\mathrm{IB} \nu} }{M_{Z'}^2}\right)\label{eq:V-mem}
\end{equation}
and is decomposed into  polarization unit vectors according to, 
\begin{equation}
    V_{\lambda_1 \lambda_2}^\mu = C_L^{\lambda_1 \lambda_2} e_L^\mu + C_R^{\lambda_1 \lambda_2} e_R^\mu + C_{s}^{\lambda_1 \lambda_2} e_{s}^\mu(\lambda_1 \lambda_2),
\end{equation} 
where the unit vectors are defined as follows:
\begin{equation}
\label{eq:pol_vectors}
\begin{split}
e_L^\mu &= \frac{1}{\sqrt{2}} \left( 0, 1, -i, 0 \right),\\
e_R^\mu &= \frac{1}{\sqrt{2}} \left( 0, -1, -i, 0 \right),\\
e_{s}^\mu(\lambda_1 \lambda_2) &= \frac{( V_{\lambda_1 \lambda_2}^0, 0, 0, V_{\lambda_1 \lambda_2}^3 )}{\sqrt{\left| \left(V_{\lambda_1 \lambda_2}^0 \right)^2 - \left(V_{\lambda_1 \lambda_2}^3 \right)^2 \right|}}.
\end{split}
\end{equation}
each polarisation component is defined as 
\begin{equation}
    C_{i}^{\lambda_1 \lambda_2} = \frac{ e_{i \mu}^{\dagger} V_{\lambda_1 \lambda_2}^\mu } { e_{i \mu}^{\dagger} e_{i}^{\mu}},
\end{equation}
where $i$ denotes the polarization (with $i = L,\, R,$ or $s$, were the last one depends on the helicities $(\lambda_1, \lambda_2)$). Notably, one finds that $e_L^\mu e_{L\mu}^\dagger = e_R^\mu e^\dagger_{R\mu} = - e_{\lambda_1 \lambda_2}^\mu e_{s\mu} (\lambda_1 \lambda_2)= -1$. 
By contracting these polarization vectors with the amplitudes, we obtain the helicity amplitudes defined as, $F_i^{V(A)} \equiv e_i^\mu F_\mu^{V(A)}$, such that $i$ is either $L$, $R$ or $s$. Consequently, the final form of the amplitude is:
\begin{widetext}
\begin{equation}
\label{eq:M_vec}
\begin{split}
\mathcal{M}(\chi(p_1, \lambda_1)& N(p_2) \to \chi(p_3, \lambda_2) N^{*} (p_4))=\\& 2W \frac{g_D g_{NZ'}}{q^2 - m^2_{Z^\prime}} \Big( \bra{N^{*}} C_L F_-^V+C_R F_+^V+C_{\lambda _1 \lambda _2} F_0^V \ket{N} - \bra{N^{*}} F_-^A C_L + F_+^A C_R + F_0^A C_{\lambda _1 \lambda _2} \ket{N} \Big).
\end{split}
\end{equation}
\end{widetext}

In \Cref{eq:V-mem}, the MEM can be  parametrized  as 
\begin{equation}
     V_{\lambda_1 \lambda_2}^{\nu }\equiv \varepsilon_{\lambda_1 \lambda_2\mu} \left(g^{\mu \nu }-\frac{q^{\mathrm{IB} \mu} q^{\mathrm{IB} \nu} }{M_{Z'}^2}\right),
\end{equation}
where the current is defined by
\begin{equation}
    \varepsilon^{\mu }_{\lambda_1 \lambda_2} \equiv \left[\bar{u}_{\chi ,\lambda _2} \gamma^{\mu }  \left(c_V^{\chi }-\gamma ^5 c_A^{\chi }\right) u_{\chi ,\lambda _1} \right]
\end{equation}
and 
in the IB frame we have:
\begin{widetext}
\begin{equation}
\label{eq:epsilon_val_IB}
\begin{split}
\varepsilon_{\lambda_1 \lambda_2}^0 &= \left( \left( 1 - 2 \delta^-_{\lambda_1} \delta^-_{\lambda_2} \right) c_V^\chi \Delta_{+,(\lambda_1 \lambda_2)} - \sigma^{\delta^-_{\lambda_1 \lambda_2}} c_A^\chi  \Delta_{-,(\lambda_1 \lambda_2)}  \right) \sqrt{1 + \lambda_1 \lambda_2 \cos \delta},\\[0.2cm]
\varepsilon_{\lambda_1 \lambda_2}^1 &= \left( 1 - 2 \delta^-_{\lambda_1} \delta^-_{\lambda_2} \right) \left( \sigma^{\delta^-_{\lambda_1 \lambda_2}} c_V^\chi \Delta_{-,(\lambda_1 \lambda_2)} - \left( 1 - 2 \delta^-_{\lambda_1} \delta^+_{\lambda_2} \right) c_A^\chi  \Delta_{+,(\lambda_1 \lambda_2)}  \right) \frac{\left||\vec{p}^{\,\mathrm{IB}}_{1}| + \lambda_1 \lambda_2 |\vec{p}^{\,\mathrm{IB}}_{3}| \right|}{|\vec{q}^{\,\,\mathrm{IB}}|} \sqrt{1 - \lambda_1 \lambda_2 \cos \delta},\\[0.2cm]
\varepsilon_{\lambda_1 \lambda_2}^2 &= i\left( \left( 1 - 2 \delta^+_{\lambda_1} \delta^-_{\lambda_2} \right) \sigma^{\delta^-_{\lambda_1 \lambda_2}} c_V^\chi \Delta_{-,(\lambda_1 \lambda_2)} -  c_A^\chi  \Delta_{+,(\lambda_1 \lambda_2)}  \right) \sqrt{1 - \lambda_1 \lambda_2 \cos \delta},\\[0.2cm]
\varepsilon_{\lambda_1 \lambda_2}^3 &= \left( 1 - 2 \delta^+_{\lambda_1} \delta^+_{\lambda_2} \right) \left( \sigma^{\delta^-_{\lambda_1 \lambda_2}} c_V^\chi \Delta_{-,(\lambda_1 \lambda_2)} - \left( 1 - 2 \delta^-_{\lambda_1} \delta^+_{\lambda_2} \right) c_A^\chi  \Delta_{+,(\lambda_1 \lambda_2)}  \right) \frac{\left||\vec{p}^{\,\mathrm{IB}}_{1}| - \lambda_1 \lambda_2 |\vec{p}^{\,\mathrm{IB}}_{3}| \right|}{|\vec{q}^{\,\,\mathrm{IB}}|} \sqrt{1 + \lambda_1 \lambda_2 \cos \delta},
\end{split}
\end{equation}
\end{widetext}
where $\sigma \equiv \mathrm{sgn}(q^2 + m_N q_0)$. We also define
\begin{equation}
    \Delta_{\alpha,\beta} \equiv \sqrt{E^\mathrm{IB}_{1}\,E^\mathrm{IB}_{3} + \alpha\, m_\chi^2 + \beta\, |\vec{p}^{\,\mathrm{IB}}_{1}|\, |\vec{p}^{\,\mathrm{IB}}_{3}|},
\end{equation}
with $\alpha$ and $\beta$ representing sign factors, and $\delta$ denoting the angle between the dark fermions in the IB frame.
The differential cross section is computed by squaring the amplitude from \Cref{eq:M_vec}, summing over all spin configurations, and then dividing by the total number of initial spin states,
\begin{widetext}
\begin{equation}
\label{eq:diff_cs_v-a}
\begin{split}
\frac{d^2 \sigma}{dq^2 dW} = \frac{g_D^2 g_{NZ'}^2}{4 \pi^2\left(q^2 - m^2_{Z^\prime}\right)^2} \frac{W}{m_N} \sum_{\lambda_1 \lambda_2} \left( \left| C_L^{\lambda_1 \lambda_2} \right|^2 \sigma_L^{\lambda_1 \lambda_2} + \left| C_R^{\lambda_1 \lambda_2} \right|^2 \sigma_R^{\lambda_1 \lambda_2} + \left| C_s^{\lambda_1 \lambda_2} \right|^2 \sigma_s^{\lambda_1 \lambda_2} \right) \delta(W-M),
\end{split}
\end{equation}
\end{widetext}
where the helicity-cross sections, $\sigma_i^{\lambda_1 \lambda_2}$, are expressed as follows:
\\
\vspace{-1cm}
\begin{equation}
\begin{split}
\sigma_R^{\lambda_1 \lambda_2} &= K \sum_{j_z} \left| \sum_{k=V,A} s_k \bra{N, j_z + 1} F_+^k \ket{N^*, j_z} \right|^2,\\[1mm]
\sigma_L^{\lambda_1 \lambda_2} &= K \sum_{j_z} \left| \sum_{k=V,A} s_k \bra{N, j_z - 1} F_-^k \ket{N^*, j_z} \right|^2,\\[1mm]
\sigma_s^{\lambda_1 \lambda_2} &= K \sum_{j_z} \left| \sum_{k=V,A} s_k \bra{N, j_z} F_{0^\pm \lambda_1 \lambda_2}^k \ket{N^*, j_z} \right|^2,
\end{split}
\end{equation}
where $K \equiv \frac{\pi}{16} \frac{W}{m_N |\vec{p}_{\mathrm{IB},1}|^2}$, and $s_V = 1$, $s_A = -1$. The helicity amplitudes are taken from the above definitions:
\begin{equation}
\begin{split}
f_{\pm |2j_z|}^{V(A)} &\equiv \bra{N, j_z \pm 1} F_\pm^{V(A)} \ket{N^*, j_z},\\[1mm]
f_{0^\pm \lambda_1 \lambda_2}^{V(A)} &\equiv \bra{N, \pm 1/2} F_{0^\pm \lambda_1 \lambda_2}^{V(A)} \ket{N^*, \pm 1/2}.
\end{split}
\end{equation}
These amplitudes will be computed using the FKR model~\cite{Feynman:1971wr}. It should be noted that we are not using the normalization for $F_0$ taken into account by Rein and Sehgal and considered in the literature~\cite{Rein:1980wg, Kuzmin:2003ji, Berger:2007rq}. To turn our amplitude into the Rein and Sehgal one, $F_0^\mathrm{RS}$, we have to do the following:
\begin{equation}
\begin{split}
F_0^\mathrm{RS} \equiv F_0 \times \sqrt{\frac{-q^2}{\left| \vec{q}^{\,\,\mathrm{IB}} \right|^2}}\,.
\end{split}
\end{equation}
%

\subsection{Differential cross section for an interaction through a scalar}

For a scalar mediated interaction, only helicity conserving transitions between 
the nucleon and the resonance occur. This is because, in the IB frame--where the 
hadron helicities are aligned--a scalar mediator cannot induce any spin flip. The amplitude takes the following form:
\begin{equation}
\label{eq:M_scalar}
\begin{split}
&\mathcal{M} (\chi(p_1, \lambda_1) N(p_2) \to \chi(p_3, \lambda_2) N^{*} (p_4))\\[0.1cm] &= \frac{g_D g_{N\Phi}}{q^2 - m^2_{\Phi}} \left[ \overline{u}_{p_3 \lambda_2} u_{p_1 \lambda_1} \right] \bra{N^{*}} J_S^{+} (0) \ket{N}, \\[0.1cm]
&= 2M \frac{g_D g_{N\Phi}}{q^2 - m^2_{\Phi}} V_S^{\lambda_1 \lambda_2} \bra{N^{*}} F_S \ket{N},
\end{split}
\end{equation}
where $J_S^{+}(0)$ is the hadronic operator for scalar interactions, and $F_S \equiv \frac{J_S^{+}(0)}{2M}$ is the scalar amplitude, defined according to the convention of normalizing the operator by twice the resonance mass $M$. In this case, the scalar MEM, defined as $V_S^{\lambda_1 \lambda_2} \equiv \left[\overline{u}_{p_3 \lambda_2} u_{p_1 \lambda_1}\right]$, is not truly a mediator that contracts with the hadronic current, since this is simply a scalar interaction. Nonetheless, we adopt the same procedure used for the vector–axial case. After working in the IB frame, its final expression is as follows:
\begin{equation}
\label{eq:V_S}
\begin{split}
&V_S^{\lambda_1 \lambda_2}\!\! = \!\left( 1 - 2 \delta^-_{\lambda_1} \delta^-_{\lambda_2} \right) \frac{\Delta^S_{(-\lambda_1 \lambda_2)}}{A_S}  \sqrt{1 \!+ \!\lambda_1 \lambda_2 \cos \delta},
\end{split}
\end{equation}
where again, $\delta$ is the angle between the incoming and outgoing $\chi$ fermions in the IB frame, the function $\Delta^S_\pm \equiv \left(E^\mathrm{IB}_1 + m_\chi\right)\left(E^\mathrm{IB}_3 + m_\chi\right) \pm |\vec{p}^{\,\,\mathrm{IB}}_1|\,|\vec{p}^{\,\,\mathrm{IB}}_3|$, and 
\begin{equation}
    A_S \equiv \sqrt{2 \left(E^\mathrm{IB}_1 + m_\chi\right)\left(E^\mathrm{IB}_3 + m_\chi\right)}. \label{eq:As_eq}
\end{equation}

The differential cross section for the process, considering the two possible helicity combinations of the hadrons ($\frac{1}{2},\frac{1}{2}$ or $-\frac{1}{2},-\frac{1}{2}$), is given by:
\begin{equation}
\label{eq:diff_cs_scalar}
\begin{split}
\frac{d^2 \sigma}{dq^2 dW} = &\frac{1}{64 \pi} \frac{g_D^2 g_{N\Phi}^2}{\left(q^2 - m^2_{\Phi}\right)^2} \frac{W^2}{m_N^2 |\vec{p}_1|^2} \\&\times \sum_{\lambda_1,\lambda_2} \left|V_S^{\lambda_1 \lambda_2} \right|^2 \sum_{i=\pm} \left|\bra{N^{*}} F_S^i \ket{N} \right|^2 \delta(W-M),
\end{split}
\end{equation}
where we have explicitly added the superscript $i$ on the amplitude, $F_S^i$, to account for the hadronic helicities in the two possible transitions:
\begin{equation}
\begin{split}
f_{0^\pm}^{S} &\equiv \langle N, \pm 1/2 | F_{0^\pm}^{S} | N^*, \pm 1/2 \rangle \,.
\end{split}
\end{equation}
where $f_{0^\pm}^{S}$ is the scalar amplitude that will be computed using the FKR model. \response{As we can see, there are only non-vanishing amplitudes for helicity conserving transitions, since the mediator is a zero-spin scalar. The same applies to pseudoscalar interactions.}

\subsection{Differential cross section for an interaction through a pseudoscalar}

We now consider a pseudoscalar-mediated interaction. As in the scalar case, only helicity-conserving transitions between the nucleon and the resonance are allowed, since in the IB frame the hadron helicities are aligned. The amplitude for this process is given by:
\begin{equation}
\label{eq:M_pseudoscalar}
\begin{split}
&\mathcal{M} (\chi(p_1, \lambda_1) N(p_2) \to \chi(p_3, \lambda_2) N^{*} (p_4))\\ &= \frac{g_D g_{Na}}{q^2 - m^2_{a}} \left[ - i \overline{u}_{p_3 \lambda_2} \gamma^5 u_{p_1 \lambda_1} \right] \bra{N^{*}} J_P^{+} (0) \ket{N}, \\[0.1cm]
&= -i2M \frac{g_D g_{Na}}{q^2 - m^2_{a}} V_P^{\lambda_1 \lambda_2} \bra{N^{*}} F_P \ket{N},
\end{split}
\end{equation}
where $J_P^{+}(0)$ is the hadronic operator for pseudoscalar interactions, and $F_P \equiv \frac{J_P^{+}(0)}{2M}$ is the pseudoscalar amplitude, normalized with respect to twice the resonance mass $M$. The pseudoscalar MEM is defined as 
$
V_P^{\lambda_1 \lambda_2} \equiv \left[\overline{u}_{p_3,\lambda_2}\,\gamma^5\,u_{p_1,\lambda_1}\right],
$
and is treated in the same manner as in the scalar-mediated case. After working in the IB frame, its final expression is as follows:

\begin{equation}
\label{eq:V_S}
\begin{split}
&V_P^{\lambda_1 \lambda_2} =( 1 - 2 \delta^-_{\lambda_1} \delta^-_{\lambda_2} )\frac{\Delta^P_{(-\lambda_1 \lambda_2)}}{A_S}  \sqrt{1 + \lambda_1 \lambda_2 \cos \delta},
\end{split}
\end{equation}
where $\delta$ is the same angle defined for the other interactions, $\Delta^P_\pm \equiv \left( E^\mathrm{IB}_3 + m_\chi \right) |\vec{p}^{\,\mathrm{IB}}_1|  \pm \left( E^\mathrm{IB}_1 + m_\chi \right) |\vec{p}^{\,\mathrm{IB}}_3|$, and $A_S$ is the same defined for the scalar mediated case defined in \Cref{eq:As_eq}.

The differential cross section for the pseudoscalar mediated interaction, considering the two possible helicity combinations of the hadrons ($\frac{1}{2},\frac{1}{2}$ or $-\frac{1}{2},-\frac{1}{2}$), is given by:
\begin{equation}
\label{eq:diff_cs_pseudoscalar}
\begin{split}
\frac{d^2 \sigma}{dq^2 dW} = \frac{1}{64 \pi} &\frac{g_D^2 g_{Na}^2}{\left(q^2 - m^2_{a}\right)^2} \frac{W^2}{m_N^2 |\vec{p}_1|^2} \sum_{\lambda_1,\lambda_2} \left|V_P^{\lambda_1 \lambda_2} \right|^2\\&\times  \sum_{i=\pm} \left|\bra{N^{*}} F_P^i \ket{N} \right|^2 \delta(W-M)\,.
\end{split}
\end{equation}

The amplitudes are defined using the same convention:
\begin{equation}
\begin{split}
f_{0^\pm}^{P} &\equiv \bra{N, \pm 1/2} F_{0^\pm}^{P} \ket{N^*, \pm 1/2} \,.
\end{split}
\end{equation}
where $f_{0^\pm}^{P}$ is the pseudoscalar amplitude that will be computed using the FKR model.

\subsection{Total cross section}
To obtain the total cross section, we integrate the differential cross section over the relevant kinematic variables. The integration limits, which are determined by the conservation of energy and momentum as well as the specific kinematics of the process, are given by:
\begin{widetext}
\begin{equation}
\begin{split}
m_N + m_\pi \le\, &\,\,W \,\,\le \sqrt{s} - m_\chi,\\[0.3cm]
2 m_{\chi }^2 - \frac{A_q B_q}{s}-\frac{\sqrt{A_q^2-s m_{\chi }^2} \sqrt{B_q^2-4 s m_{\chi }^2}}{s} \le \,&\,\,\,|q^2| \,\le 2 
m_{\chi }^2 - \frac{A_q B_q}{s}+\frac{\sqrt{A_q^2-s m_{\chi }^2} \sqrt{B_q^2-4 s m_{\chi }^2}}{s},
\end{split}
\end{equation}
\end{widetext}
where $A_q \equiv \left(E_1 m_N+m_{\chi }^2\right)$, $B_q \equiv \left(m_{\chi }^2+s-W^2\right)$, and $s$ is the Mandelstam variable. 

The expressions obtained in this section assume a narrow width for the produced resonance—i.e., that it is completely on-shell, as indicated by the delta distribution $\delta(W-M)$. To account for resonances with a finite width, this delta function must be replaced by a Breit–Wigner factor:
\begin{equation}
\label{eq:Breit-Wigner}
\delta(W - M_r) \to |\eta_r (W)|^2.
\end{equation}
The parameter $\eta_r(W)$ must be multiplied with the amplitude before squaring it. It is defined as:\footnote{Note that in \Cref{eq:eta} the normalization factor, $N_r$, will not be used for the vector-axial case since we use the dipole functions modelled by Ref.~\cite{Graczyk:2007bc}, who employed the Rarita-Schwinger formalism, which uses a non-normalized Breit-Wigner factor~\cite{smith1972neutrino}.}
\begin{equation}
\label{eq:eta}
\eta_r(W) = \sqrt{\frac{\Gamma_r}{2\pi N_r}}\,\frac{1}{W - M_r + i\Gamma_r/2},
\end{equation}
where we have now added the subscript $r$ to explicitly indicate that this must be done for each resonance and $N_r$ is the normalization factor equal to:
\begin{equation}
\begin{split}
N_r = \int_{W_\mathrm{min}}^\infty dW \frac{\Gamma_r}{2 \pi} \frac{1}{(W - M_r)^2 + \Gamma_r^2/4}\,.
\end{split}
\end{equation}
Following the method described in Ref.~\cite{Rein:1980wg}, the decay width $\Gamma_r$ is computed as
\begin{equation}
\Gamma_r = \Gamma_r^0 \left( \frac{q_\pi(W)}{q_\pi(M_r)} \right)^{2L+1},
\end{equation}
with the values of $\Gamma_r^0$ provided therein, $L$ denoting the orbital angular momentum of the resonance, and
\begin{equation}
q_\pi(x) = \frac{\sqrt{(x^2 - m_N^2 - m_\pi^2)^2 - 4m_N^2 m_\pi^2}}{2x}.
\end{equation}

Following the approach in Ref.~\cite{Rein:1980wg}, the amplitude must be multiplied by the square root of the elasticity, $\sqrt{x_E^r}$, to incorporate the branching ratio into the final state ($N' \pi$) in the computation. Additionally, the sign of the decay amplitude of the resonance, $\mathrm{sgn} (N^*_r)$, must be included to ensure a correct summation of the processes. Furthermore, the isospin of the final state ($N' \pi$) must be accounted for, with its weight given by the Clebsch–Gordan coefficient, $a_\mathrm{CG}$. The values for different cases, as listed in Section 2.2(a) of Ref.~\cite{Rein:1980wg}, are $\sqrt{2/3}$ ($\sqrt{1/3}$) for $\Delta$ resonances decaying into $\pi^0$ ($\pi^\pm$), while for $N$ resonances, the coefficients are reversed: $\sqrt{1/3}$ ($\sqrt{2/3}$) for $\pi^0$ ($\pi^\pm$), with a negative (positive) sign when the outgoing particle is a proton (neutron).

In summary, the amplitude transformation required is:
\begin{equation}
\mathcal{M} \to a_\mathrm{CG} \times \mathcal{M} \times \eta_r(W) \times \sqrt{x_E^r} \times \mathrm{sgn} (N^*_r).
\end{equation}
The amplitudes for each resonance can be summed coherently if the total and orbital angular momenta are the same, so summing the different isospin ($3/2$ and $1/2$) cases before squaring the amplitude. For further details on this computation, we refer the reader to Section 2.2 of Ref.~\cite{Rein:1980wg}. For the angular distribution of the products of the interactions, and particularly the pions, we refer the reader to~\cite{Rein:1987}. The angular formalism is not directly affected by the new physics and, therefore, it is out of the scope of the present study.

\begin{figure}
    \includegraphics[width=0.48\textwidth]{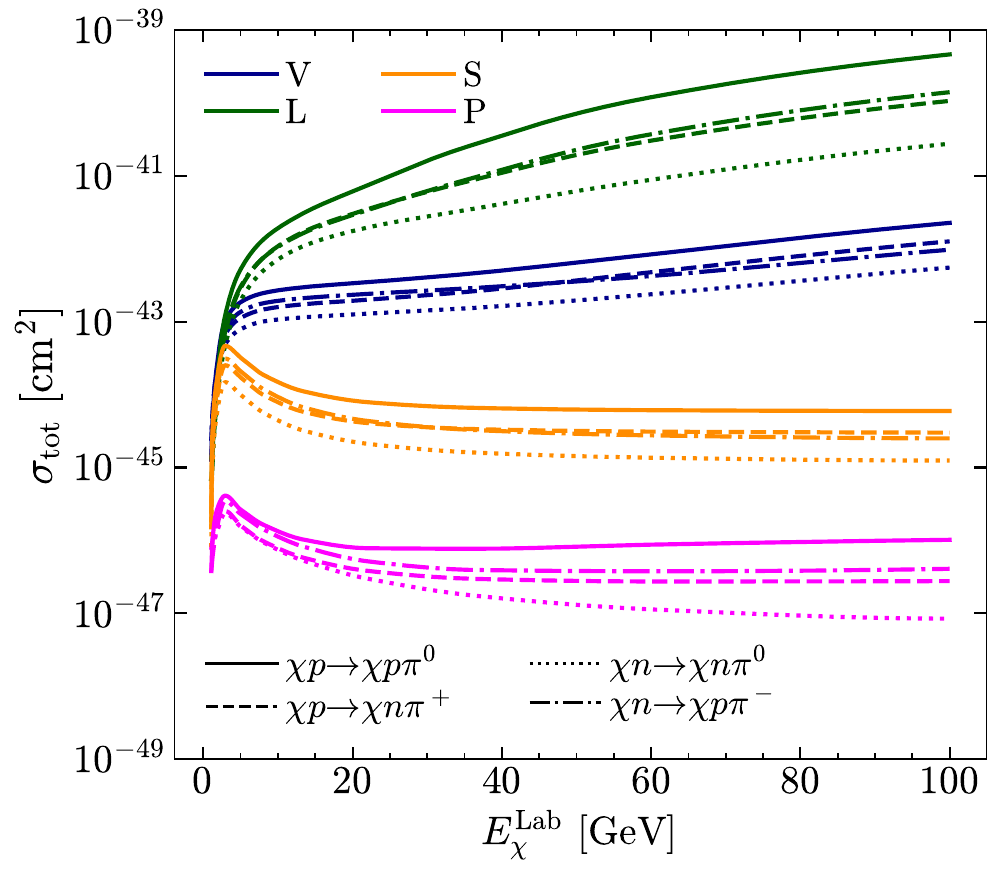}
    \caption{ Cross sections of the four channels that produce single pions in the final state through a resonance for four cases: dark photon mediator with just vector (V) or pure left (L) couplings, scalar (S), and pseudoscalar (P) mediators. We have set the masses to $m_{Z^\prime,\phi,a} = 1\,\rm GeV$ and $m_\chi = 0.5\, \rm GeV$, while the couplings are set to $g_D = 0.1$, $g_{NZ^\prime,\phi} = 10^{-6}$ and $g_{Na}=10^{-5}$. We also set $g_u^V = 2 g_d^V = 2$ and $g_u^A = g_d^A = 0$ for the vector case, while $g_u^V = 2g_d^V = 1$ and $g_u^A = 2g_d^A = 1$ for the left case.
     \label{fig:cross_sections}}
\end{figure}

Just as an example of the cross sections obtained, in \Cref{fig:cross_sections}, we present the total cross section for four different interaction cases: vector (V), left (L), scalar (S), and pseudoscalar (P), including all possible channels. We fix the mass of the mediator for each case to $m_{Z^\prime, \phi, a} = 1, \rm GeV$ and the mass of the dark fermion to $m_\chi = 0.5, \rm GeV$. The couplings to the Standard Model are set to $g_{N Z^\prime, \phi} = 10^{-6}$ for the vector-axial and scalar cases, while for the pseudoscalar case, it is fixed to $g_{N a} = 10^{-5}$. We also set $g_u^V = 2 g_d^V = 2$ and $g_u^A = g_d^A = 0$ for the vector case, while $g_u^V = 2g_d^V = 1$ and $g_u^A = 2g_d^A = 1$ for the left case. The coupling to the dark sector is set to $g_D = 0.1$. The pseudoscalar coupling is set to a higher value because its cross section is suppressed compared to the other cases. Since the plots are only meant to illustrate the results, this choice does not affect the conclusions drawn from them.

For the vector-axial cases, the cross sections increase as the energy increases. However, at even higher energies, the cross sections begin to be  suppressed due to the limited parameter space available to produce the resonance nearly on-shell. The scalar and pseudoscalar cases exhibit a very similar pattern: both show a peak around $3$ GeV, followed by a stabilization of the cross section, which is also  suppressed at higher energies than those shown in \Cref{fig:cross_sections}.

There is a twofold effect coming from a pion pole that affects both the axial and the pseudoscalar amplitudes, as we will discuss in \Cref{sec:form-factors}. This inclusion is dictated by the partial conservation of the axial current (PCAC). In the case of the axial amplitude, there is an additional factor given by $\eta^{\mu \nu} + q^\mu q^\nu / \left(m_\pi^2 - q^2 \right)$, such that $\eta^{\mu \nu}$ is the Minkowski metric, which strengthens the interaction at higher energies. This is why the left (L) case exhibits higher cross sections, coming from the axial contribution. For the pseudoscalar case, the extra factor is $m_\pi^2 / \left(m_\pi^2 - q^2 \right)$, which for the values of $q^2$ that are relevant in our parameter space,  diminishes the values of the amplitudes, leading to lower cross sections.

\section{Amplitudes}
\label{sec:form-factors}

In this section, we describe the FKR model, used to calculate the amplitudes, and explain how to perform the computation for different mediators: vector, axial, scalar, and pseudoscalar bosons. The explicit list of amplitudes, obtained by considering the transitions from a proton to a resonance mediated by the operators presented below, is provided in \Cref{tab:form_factors} of \Cref{sec:table_form_factors}. To obtain the transitions from neutrons, we simply interchange the up quark couplings with those for down quarks.

\subsection{FKR model}
The FKR relativistic quark model provides an effective framework for describing hadronic interactions in the resonance region~\cite{Feynman:1971wr}. In this model, hadrons are treated as excitations of a four-dimensional harmonic oscillator, which enables relatively simple calculations of transition amplitudes. By prioritizing simplicity over certain theoretical details, the FKR approach achieves computational efficiency while still yielding results that are in good agreement with experimental observations, even though some corrections are necessary to fully account for the observations. This balance of simplicity and certain degree of accuracy makes the model particularly useful for applications within the Standard Model, and it has become a valuable tool in modern neutrino event generators like GENIE~\cite{Andreopoulos:2009rq}.

The theory is based on three constants used to fit the experimental data: (1) the the level spacing per unit of angular momentum along the trajectories, $\Omega = 1.05 \ \mathrm{GeV}^2$, assumed to be identical for mesons and baryons; (2) a parameter that quantifies the strength of the pseudoscalar-meson coupling to hadrons, $f_R = 1.21 \ \mathrm{GeV}^{-1}$; and (3) an adjustment factor $F$, implemented as a Gaussian function of the center-of-mass momentum, which serves to reduce  discrepancies between the predictions and data at higher resonance masses.

The model organizes baryons into   SU($6$) multiplets, namely the  \underline{$56$}, \underline{$70$} and \underline{$20$} representations. 
These multiplets originate from the tensor product of the spin, SU($2$), and the flavor (isospin), SU($3$) groups for three quarks. 
Additionally, the model incorporates excited states of two three-dimensional modes of internal harmonic oscillation among the quarks. 
Baryonic states are denoted as $\left[ \underline{A}, L^P \right]_N$, where $\underline{A}$ is one of the multiplets (\underline{$56$}, \underline{$70$} or \underline{$20$}), $L$ is the total orbital angular momentum, $N$ is the number of oscillator quanta in the  resonance, and $P = (-1)^N$ is the parity. 

The SU($3$) multiplet within each group is further specified as $^d \left(\underline{B} \right)_J$, where $\underline{B}$ represents the SU($3$) multiplet-- \underline{$1$} for singlets (e.g. $\Lambda$), \underline{$8$} for octets (e.g. $N$), or \underline{$10$} for decimets (e.g. $\Delta$); $d$ denotes the spin multiplicity ($4$ for spin-$\frac{3}{2}$ and $2$ for spin-$\frac{1}{2}$); and $J$ is the total angular momentum of the state. Notably, no baryon state is found in the  \underline{$20$} representation. For our purposes, we focus solely on the proton, the neutron and the 18 resonances considered in the analysis by, Rein and Sehgal~\cite{Rein:1980wg}, without addressing meson classification.

The model is formulated as a harmonic oscillator whose eigenvalues are given by $m^2$ and vary as $N \Omega$, where $N$ is the number of oscillator quanta. The Hamiltonian is expressed as,
\begin{equation}
\begin{split}
\mathcal{H} = &3 \sum_k p_k^2 + \frac{1}{36} \Omega^2 \sum_{j<k} \left(\Delta u_{jk}\right)^2 + \mathrm{const.},
\end{split}
\end{equation}
with the indices  $j$ and $k$ running over the three quarks, $p_k$ denoting the four-momentum operator of  the $k$th quark, and where the difference $\Delta u_{jk} \equiv (u_j - u_k)$ involves the position operator $u_k$, which is  conjugate to the momentum $p_{k \mu} = i (\partial/\partial u_k^\mu)$. By introducing center-of-mass and relative coordinates, see Equations (4a) and (4b) in \cite{Feynman:1971wr},  and subsequently defining creation and annihilation operators as in Equations (12a) and (12b) of the same reference, one obtains a simplified version of the Hamiltonian:
\begin{equation}
\label{eq:Ham_FKR}
\begin{split}
\mathcal{H} = &P^2 - \Omega \left( a_\mu^* a^\mu + b_\mu^* b^\mu \right) + \mathrm{const}.
\end{split}
\end{equation}
Here $a(^*)$ and $b(^*)$ denote two sets of  annihilation (creation) operators and $P = \sum_k p_k$ is the total momentum of the three quark system. A system of three quarks has only two independent relative coordinates, which are represented by these two types of operators. They satisfy the algebra $\left[ a_\mu, a_\nu^* \right] = \left[ b_\mu, b_\nu^* \right] = - \eta_{\mu \nu}$, where $\eta$ is the Minkowski metric.  Assuming that the interactions are symmetric for all quarks, we focus solely on the $a$ excitations. These operators generate states with quantum number $N$ from states with quantum number $N-1$ states. For $N=1$, \vspace{-0.5cm}
\begin{equation}
\label{eq:FKR_a_operators}
\begin{split}
\ket{1,+1} &= a_+^\dagger \ket{g},\\
\ket{1,0} &= a_z^\dagger \ket{g},\\
\ket{1,-1} &= a_-^\dagger \ket{g},
\end{split}
\end{equation}
where,
\begin{equation}
    a^\dagger_\pm \equiv \mp \frac{\left( a^\dagger_x \pm i a^\dagger_y \right)}{\sqrt{2}},
\end{equation}
with $\ket{g}$ representing the ground state ($N=0$), and the states are labeled as $\ket{N, L}$. The states created by $a^\dagger$ exhibit an $\alpha$-symmetry (for further details,  we refer the reader to the first appendix of \cite{Feynman:1971wr}).
The non-kinetic part of the Hamiltonian in \Cref{eq:Ham_FKR} has as eigenvalues that are integer multiples of $\Omega$. This  Hamiltonian form the basis for deriving the interaction terms that are used to construct the transition amplitudes between baryonic states. 

\response{The group structure in the Hamiltonian also allows to extract the couplings that appear for each transition in terms of the quark couplings. This structure depends on the group to which the particles involved belong to: the octet or the decimet, and also on the symmetry structure of the wave function ($S$, $\alpha$, $\beta$, $A$), as can be seen in Eq. (A12) in \cite{Feynman:1971wr}. For interactions with protons, if $e_k$ represents the charge or coupling operator for the transition, the only non-vanishing couplings are: 
\begin{equation}
\begin{split}
\bra{(8)_\alpha} e_k \ket{(10)_S} &= \frac{\sqrt{2}}{3} \left( c_u - c_d \right)\,,\\
\bra{(8)_\alpha} e_k \ket{(8)_\alpha} &= \frac{1}{3} \left( c_u + 2 c_d \right)\,,\\
\bra{(8)_\beta} e_k \ket{(8)_\beta} &= c_u\,,
\end{split}
\end{equation}
where $c_q$ is the couplings with the quark $q$. We will parametrize them as $c_q \equiv g_{N(Z',\Phi, a)} g_q$, so that an overall factor $g_{N(Z',\Phi, a)}$ gives us the order of magnitude of all the couplings for each interaction (through a dark photon, a scalar or a pseudoscalar) For transitions from neutrons, we just replace $c_u \leftrightarrow c_d$.
} \response{This structures can also make some transitions to vanish, depending on the coefficient arrange of the couplings to quarks. For example, equal couplings to the up and down quark would directly translate into a suppressed transition from $\ket{(8)_\alpha}$ states to $\ket{(10)_S}$.}

\subsection{Vector mediated interactions}

To account for the spin structure of the quarks, the FKR model interprets the $p_k^2$ terms in \Cref{eq:Ham_FKR} as $\slashed{p}_k \slashed{p}_k$. 
The  vectorial interaction for a vector boson $A^\mu$ is then obtained by replacing $\slashed{p}_k \to \slashed{p}_k - e_k \slashed{A} (u_k)$, where $e_k$ denotes the unitary spin matrix for the $k$th quark.

Expanding the product  $( \slashed{p}_k - e_k \slashed{A} (u_k) )^2$ up to first order in $e_k$ yields the interaction term,
\begin{equation}
\begin{split}
    \label{eq:V_int_FKR_0}
\mathcal{M}^V &= e_\mu \left( 3 \sum_k e_k \left( \slashed{p}_k \gamma^\mu e^{i q u_k} + \gamma^\mu e^{i q u_k} \slashed{p}_k \right) \right)\\
&\equiv 2 W \left( e^\mu F_\mu^V \right).
\end{split}
\end{equation}
In this expression,  $q$ is the momentum of the vector boson, $e_\mu$ is its polarization vector,  and $F_\mu^V$ is the vector amplitude introduced in \Cref{eq:M_vec}.  By evaluating the knematical variables  in the IB frame and accounting for the longitudinal polarization coming from the MEM, one obtains:
\begin{widetext}
\begin{equation}
\begin{split}
\mathcal{M}^V= &9 F^V (q^2)\, e_a\, \mathrm{exp}\left(\!{-\sqrt{\frac{2}{\Omega}} q^\mathrm{IB} \cdot a^\dagger}\right)\Bigg\{ \!\!\bigg[ \frac{2}{3} M \!-\! \frac{1}{3} q_0^\mathrm{IB} -\! \frac{|\vec{q}^{\,\,\mathrm{IB}}|^2}{2 m_N  g^2} - \frac{2}{3} \sqrt{\frac{\Omega}{2}} \left( a_t + a_t^\dagger \right) \bigg] e_t\\
&+ \frac{2}{3} \sqrt{\frac{\Omega}{2}} \left( \vec{a} + \vec{a}^{\,\dagger} \right) \cdot \vec{e} + \left( \vec{q}^{\,\,\mathrm{IB}}\cdot \vec{e} \right) \left(\frac{1}{3} + \frac{|q_0^\mathrm{IB}|}{2 m_N  g^2} \right) 
 + i \vec{\sigma} \cdot \left( \vec{q}^{\,\,\mathrm{IB}}\times \vec{e} \right) \!\!\left(1 + \frac{|q_0^\mathrm{IB}}{2 m  g^2} \right)  \!\!\!\Bigg\} \mathrm{exp}\left(\!\sqrt{\frac{2}{\Omega}} q^\mathrm{IB} \cdot a\right)
 \end{split}
 \label{eq:V_int_FKR}
\end{equation}
\end{widetext}
where the parameter $g^2$  is defined by 
\begin{equation}
\label{eq:g_deff}
    g^2\equiv \frac{(m_N+W)^2-q^2}{4 W m_N}  
\end{equation}
and $\vec{\sigma}$ are the Pauli matrices.  In this derivation, we consider that  each quark contributes identically,  which is why a common factor  $e_a$ is introduced. Moreover, a  factor of $e^{q^2/\Omega} \times g^3$ is replaced by $F^V (q^2)$,  which is modeled as a modified dipole factor.

Finally, when  the polarization vectors (as defined in \Cref{eq:pol_vectors}) are taken into account, the transition operators are obtained as,
\begin{equation}
\label{eq:FKR_vector_Fs}
\begin{split}
&F^V_\pm = -9 e_a \left( R^V \sigma_\pm + T^V a_\mp \right) e^{-\lambda a_z},\\
&F^{V,{\lambda_1 \lambda_2}}_{0^\pm} = 9 e_a S_{\lambda_1 \lambda_2} e^{-\lambda a_z},
\end{split}
\end{equation}
with the parameters defined by

\begin{equation}
\label{eq:FKR_vector_params}
\begin{split}
\lambda &\equiv \sqrt{\frac{2}{\Omega}} \frac{m_N}{W} |\vec{q}|,\\
R^V &\equiv \frac{\sqrt{2} m_N}{W} \frac{\left(W + m_N \right) |\vec{q}|}{\left(W + m_N \right)^2 - q^2} F^V (q^2),\\
T^V &\equiv \frac{\sqrt{\Omega} }{3 \sqrt{2} W} F^V (q^2),\\
S_{\lambda_1 \lambda_2} &\equiv \frac{m_N}{W \left|\vec{q}\right|} \frac{V_{\lambda_1 \lambda_2}^3 q^0_\mathrm{IB} - V_{\lambda_1 \lambda_2}^0 \left|\vec{q}_\mathrm{IB}\right|}{C_s^{\lambda_1 \lambda_2}} \\
& \ \ \ \times \left(\frac{1}{6} - \frac{q^2}{6 m_N^2} - \frac{W}{2m_N} \right) F^V (q^2)\,.
\end{split}
\end{equation}
In Ref.~\cite{Graczyk:2007xk}, $F^V (q^2)$ is modeled as: 
\begin{equation}
\label{eq:F_V}
\begin{split}
F^V (q^2) = &\frac{1}{2} \sqrt{1 - \frac{q^2}{(m_N+W)^2}} \left(1 - \frac{q^2}{4 m_N^2} \right)^{-N} \\[0.1 cm]& \times \sqrt{3 \left( G_3 (q^2) \right)^2 + \left( G_1 (q^2) \right)^2}
\end{split}
\end{equation}
\\
where the functions $G_1 (q^2)$ and $G_3 (q^2)$ are defined in Equations (38-39) of \cite{Graczyk:2007bc}, and $N$ is the total quantum number.

To obtain the amplitude for each transition, one must to consider the different resonances and their representations in the SU($6$) group.

\subsection{Axial mediated interactions}

In the case of an axial coupling, the axial boson $B^\mu$ is introduced by replacing  $\slashed{p}_k \to \slashed{p}_k - e_k \gamma^5\slashed{B} (u_k)$. Expanding the product$\left( \slashed{p}_k - e_k \gamma^5 \slashed{B} (u_k) \right)^2$ to first order in $e_k$ in \Cref{eq:Ham_FKR}, leads to the interaction term
\begin{equation}
\begin{split}
\label{eq:A_int_FKR_0}
\!\!\!\!\!Z^{-1} \mathcal{M}^A&=  3i e_\mu \! \sum_k e_k \left( \slashed{p}_k \gamma^5 \gamma^\mu e^{i q u_k} + \gamma^5 \gamma^\mu e^{i q u_k} \slashed{p}_k \right) \\
& \equiv 2 W \left( e^\mu F_\mu^A \right)
\end{split}
\end{equation}
Here,  $q$ is the boson momentum, $e_\mu$ is its polarization vector, and $F_\mu^A$ is the axial amplitude introduced in \Cref{eq:M_vec}. The factor $Z$ is a renormalization constant required because  the axial current is not conserved;  the FKR model predicts an axial vector amplitude at  $q^2 = 0$ equal to $5/3$, while  the experimental value is $5/4$, hence $Z = 3/4$.

When the parameters are  expressed in the IB frame,  the amplitude becomes\vspace{5cm}
\begin{widetext}
    \begin{equation}
\label{eq:A_int_FKR}
\begin{split}
\!\!\!\!Z^{-1}\! \mathcal{M}^A\! =&\! -9 F^A (q^2) e_a \mathrm{exp}\left(\!{\!-\sqrt{\frac{2}{\Omega}} q^\mathrm{IB} \cdot a^\dagger}\right)\!\!\Bigg\{ \!\!\big(\vec{\sigma} \cdot \vec{q}^{\,\,\mathrm{IB}}\big)\!\bigg[\frac{1}{3} e_t\! 
+\! \frac{q_0^\mathrm{IB} e_t}{2 m_N g^2} \!-\! \frac{2}{3} \frac{\vec{q}^{\,\,\mathrm{IB}}\cdot \vec{e}}{2 m_N g^2} \bigg] \!\!+\!\! \frac{2}{3} \sqrt{\frac{\Omega}{2}} \vec{\sigma} \cdot ( \vec{a}\! +\! \vec{a}^{\,\,\dagger}) \!\! \left[e_t + \frac{\vec{q}^{\,\,\mathrm{IB}}\cdot \vec{e}}{2 m_N g^2} \right]\!\\[0.2cm]
&+\! (\vec{\sigma} \cdot \vec{e} ) \!\bigg[ \frac{2}{3} M \!-\! \frac{1}{3} q_0^\mathrm{IB}\! -\! \frac{1}{3} \frac{|\vec{q}^{\,\mathrm{IB}}|^2}{2 m_N  g^2} \!-\! \frac{2}{3} \sqrt{\frac{\Omega}{2}} \frac{\vec{q}^{\,\,\mathrm{IB}}\cdot \left( \vec{a} \!+\! \vec{a}^{\,\,\dagger} \right)}{2 m_N g^2} \bigg]\!+\! i\frac{2}{3} \sqrt{\frac{\Omega}{2}} ( \vec{a} + \vec{a}^{\,\,\dagger} ) \!\cdot \!\frac{\vec{q}^{\,\,\mathrm{IB}}\times \vec{e}}{2 m_N g^2}  \Bigg\} \mathrm{exp}\left(\!\sqrt{\frac{2}{\Omega}} q^\mathrm{IB} \cdot a\right)
\end{split}
\end{equation}
\end{widetext}
$g$ is defined in \Cref{eq:g_deff}, and a common factor $e_a$ in included since each quark contributes identically. Additionally, a factor   $e^{q^2/\Omega} \times g^3$  is absorbed into  $F^A (q^2)$, which is modeled similarly to the vector case. Upon incorporating the polarization vectors, as defined in  \Cref{eq:pol_vectors}, the transition operators are obtained, 
\begin{equation}
\label{eq:FKR_axial_Fs}
\begin{split}
F^A_\pm &= -9 e_a \left( R^A \sigma_\pm + T^A a_\mp \right) e^{-\lambda a_z},\\[0.1cm]
F^{A,{\lambda_1 \lambda_2}}_{0^\pm} &= - 9 e_a \left[ C_{\lambda_1 \lambda_2} \sigma_z + B_{\lambda_1 \lambda_2} \vec{\sigma} \cdot \vec{a} \right] e^{-\lambda a_z},
\end{split}
\end{equation}
with the parameters defined%
\begin{align*}
R^A \!\equiv& \frac{\sqrt{2}Z}{6 W}\! \left(W \!+\! m_N \!+ \!\frac{2 N \Omega W}{(W + m_N)^2 - q^2 } \right)\!  F^A (q^2),\\[0.1cm]
T^A \!\equiv &\frac{\sqrt{2\Omega} Z}{3} \frac{m_N |\vec{q}|}{W \left((W + m_N)^2 - q^2 \right)} F^A (q^2),\\[0.1cm]
B_{\lambda_1 \lambda_2} \!\equiv &\frac{1}{C_s^{\lambda_1 \lambda_2}} \sqrt{\frac{\Omega}{2}}\! \left(V_{\lambda_1 \lambda_2}^0 \!\!+\! V_{\lambda_1 \lambda_2}^3 \frac{|\vec{q}^{\,\mathrm{IB}}|}{a m_N} \right)\! Z \frac{F^A (q^2)}{3 W},\nonumber
\end{align*}
\begin{equation}
\label{eq:FKR_axial_params}
\begin{split}
C_{\lambda_1 \lambda_2} \!\equiv& \frac{1}{C_s^{\lambda_1 \lambda_2}} \!\bigg[ ( V_{\lambda_1 \lambda_2}^0 |\vec{q}^{\,\mathrm{IB}}|\! -\! V_{\lambda_1 \lambda_2}^3 q_0^\mathrm{IB} )\!\! \left(\frac{1}{3} \!+\! \frac{q_0^\mathrm{IB}}{a m_N} \right) \\[0.1cm]
  +& V_{\lambda_1 \lambda_2}^3 \left( \frac{2}{3} W + \frac{q^2}{a m_N} + \frac{N \Omega}{3 a m_N} \right) \bigg] \frac{Z}{2 W} F^A (q^2),\\[0.1cm]
 a \equiv &1 + \frac{W^2 + m_N^2 - q^2}{2 m_N W}.
\end{split}
\end{equation}
Here $N$ denotes the total quantum number. These expressions account for massive non-hadronic currents (such as those from dark fermions) and were initially derived in Ref.~\cite{Kuzmin:2003ji}. Later, Ref.~\cite{Berger:2007rq} introduced corrections to two terms due to the pion-pole contribution in the hadronic axial current (as dictated by PCAC). These corrections modify the parameters as follows:
\begin{equation}
\label{eq:FKR_axial_params_PCAC}
\begin{split}
B_{\lambda_1 \lambda_2}\!\!\! \to &B_{\lambda_1 \lambda_2}\! +\! \frac{1}{C_s^{\lambda_1 \lambda_2}} \frac{Z}{2 W } \left( V_{\lambda_1 \lambda_2}^0 q_0^\mathrm{IB} \!-\! V_{\lambda_1 \lambda_2}^3 |\vec{q}^{\,\mathrm{IB}}| \right) \\[0.1cm]
&\times \frac{\frac{2}{3}\sqrt{\frac{\Omega}{2}} \left( q_0^\mathrm{IB} + \frac{|\vec{q}^{\,\mathrm{IB}}|^2}{a m_N} \right)}{m_\pi^2 - q^2} F^A (q^2),\\[0.1cm]
 C_{\lambda_1 \lambda_2}\!\!\! \to &C_{\lambda_1 \lambda_2} \!+\!  \frac{1}{C_s^{\lambda_1 \lambda_2}} \frac{Z}{2 W}  \left( V_{\lambda_1 \lambda_2}^0 q_0^\mathrm{IB} \!-\! V_{\lambda_1 \lambda_2}^3 |\vec{q}^{\,\mathrm{IB}}| \right) \\[0.1cm]
 &\times \frac{|\vec{q}^{\,\mathrm{IB}}| \left( \frac{2}{3} W + \frac{q^2}{a m_N} + \frac{N \Omega}{3 a m_N} \right)}{m_\pi^2 - q^2}  F^A (q^2)\,.
\end{split}
\end{equation}

Furthermore, in Ref.~\cite{Graczyk:2007xk}, the amplitude $F^A (q^2)$ is modeled as:
\begin{equation}
\label{eq:F_A}
\begin{split}
F^A (q^2) = &\frac{\sqrt{3}}{2} \sqrt{1 - \frac{q^2}{(m_N+W)^2}} \left(1 - \frac{q^2}{4 m_N^2} \right)^{-N} \\[0.1cm]
& \times \left[1 - \frac{W^2 + q^2 - m_N^2}{8 m_N^2} \right] C_5^A (q^2)\,
\end{split}
\end{equation}
with $C_5^A (q^2)$ defined in Equations (52-53). We adopt the first definition provided there. 

\subsection{Scalar mediated interactions}

Since the FKR model does not include scalar interactions with quarks, we introduce an additional Yukawa scalar term following the procedure in Ref.~\cite{Feynman:1971wr}. This term is given by:
\begin{equation}
\label{eq:scalar_int_FKR_0}
\begin{split}
\mathcal{M}^S &= 2 W \sum_k e_k e^{i q \cdot u_k},\\
&\equiv 2 W F^S
\end{split}
\end{equation}
where $F^S$ is the scalar amplitude introduced in Eq.~(\ref{eq:M_scalar}), and the factor $2W$ is included to maintain consistency with the conventions used in the vector and axial cases. In the FKR approach, the transition matrix element is modeled from $2mH$, where $H$ is the interaction Hamiltonian and $m$ represents the mass of the harmonic oscillators; we reintroduce this factor here to facilitate its separation from the amplitude in Eq.~(\ref{eq:M_scalar}).

When expressed in the IB frame, the scalar amplitude becomes:
\begin{equation}
\label{eq:FKR_scalar_Fs}
\begin{split}
F^{S}_{0^\pm} = e_k\,S^S\,e^{-\lambda a_z},
\end{split}
\end{equation}
with
\begin{equation}
\label{eq:FKR_scalar_params}
\begin{split}
S^S \equiv 3\,F^S(q^2).
\end{split}
\end{equation}
The function $F^S(q^2)$ is modeled as a scalar dipole factor. As a reasonable approximation, we adopt the pion amplitude fitted by Ref.~\cite{Alexandrou:2021ztx}:
\begin{equation}
F^S(q^2) = \frac{F^S(0)}{1 - \frac{q^2}{m_S^2}},
\end{equation}
with $F^S(0) = 1.165$ and $m_S = 1.221\,\mathrm{GeV}$. \response{The validity of this choice could only be tested in future BSM measurements, if any. The same will apply for the pseudoscalar case.}

\subsection{Pseudoscalar mediated interactions}
\label{sec:pseudoscalar_form_factors}

The FKR model originally considered a pseudoscalar interaction (see Equation (42) in Ref.~\cite{Feynman:1971wr}) by replacing the polarization vector $e_\mu$ in the axial interaction (Eq.~\ref{eq:A_int_FKR}) with $-i\,q_\mu$ to model decays of hadrons into another baryon and a pseudoscalar meson. However, this approach is more appropriate for meson interactions. Instead, we adopt an alternative method—also mentioned in Ref.~\cite{Feynman:1971wr}—by directly including a $\gamma^5$ term that is independent of the axial interaction. The resulting interaction is given by:
\begin{equation}
\label{eq:pseudoscalar_int_FKR_0}
\begin{split}
\mathcal{M}^P &= 2i\,W\,\sum_k e_k\,\gamma^5\,e^{i\,q\cdot u_k},\\[0.1cm]
&\equiv 2W\,i\,F^P,
\end{split}
\end{equation}
where $F^P$ is the scalar amplitude introduced in Eq.~(\ref{eq:M_pseudoscalar}). Upon introducing the IB frame parameters, the pseudoscalar amplitude is obtained as:
\begin{equation}
\label{eq:FKR_pseudscalar_Fs}
\begin{split}
F^P_{0^\pm} = e_k\,S^P\,\sigma_z\,e^{-\lambda a_z},
\end{split}
\end{equation}
with
\begin{equation}
\label{eq:FKR_pseudoscalar_params}
\begin{split}
S^P \equiv \frac{6\,m_N\,|\vec{q}|}{(W+m_N)^2-q^2}\,F^P(q^2).
\end{split}
\end{equation}
The function $F^P(q^2)$ is modeled as a pseudoscalar dipole factor:
\begin{equation}
F^P(q^2) = \frac{F^P(0)}{1-\frac{q^2}{m_P^2}},
\end{equation}
where the fitted parameters $F^P(0) = 1.25$ and $m_P = 1.23 \ m_N$ are taken from \cite{Chen:2021guo}. They are used for an axial dipole modeling, though due to its relation with pseudoscalar the amplitude, we take it as a test value.

\begin{figure*}
\centering
    \includegraphics[width=0.49\textwidth]{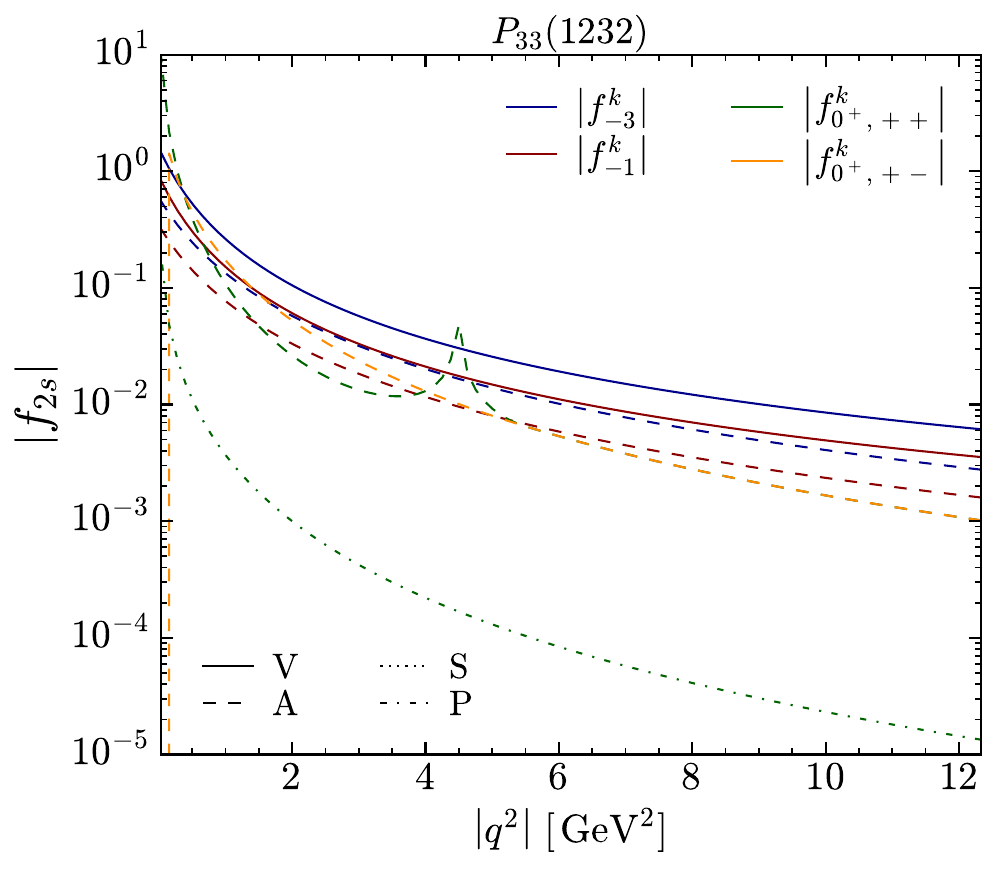}
    \includegraphics[width=0.49\textwidth]{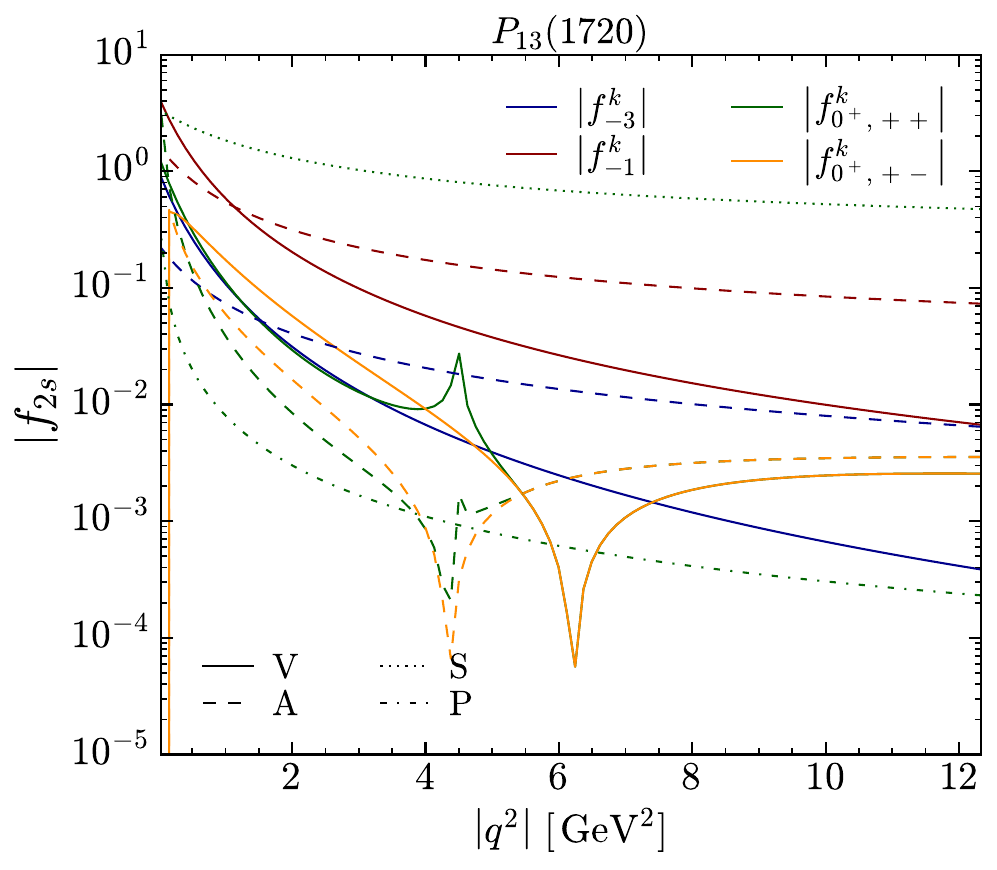}
    \caption{ Amplitudes for the transition from a proton to a resonance: $P_{33} \ (1232)$ (left panel) and $P_{13} \ (1720)$ (right panel). We can see the vectorial (solid), axial (dashed), scalar (dotted) and pseudoscalar (dashed-dotted) amplitudes for an incoming dark fermion of $10$ GeV of energy.
     \label{fig:form_factors}}
\end{figure*}
As with the axial case, the pseudoscalar amplitude is modified by a pion-pole factor required by PCAC. To account for this effect, consider the pseudoscalar matrix element generated from an axial current,
\[
M^{A\to P} \equiv M^A\Bigl(e^\mu\to -i\,q^\mu\Bigr),
\]
where $M^A = e^\mu J^A_\mu$ and $J^A_\mu$ is the axial current. According to Ref.~\cite{Berger:2007rq}, the axial current transforms as
\begin{equation}
J^A_\mu \to J^A_\mu + q_\mu\,\frac{q^\nu J^A_\nu}{m_\pi^2 - q^2}.
\end{equation}

%
Applying the same transformation to $M^{A\to P}$ yields
\begin{equation}
\label{eq:pseudoscalar_PCAC}
\begin{split}
M^{A\to P} \to M^{A\to P}\,\left(\frac{m_\pi^2}{m_\pi^2-q^2}\right).
\end{split}
\end{equation}
Thus, assuming that our amplitude behaves the same way, the parameter $S^P$ is modified to
\begin{equation}
\begin{split}
S^P &\to S^P\,\left(\frac{m_\pi^2}{m_\pi^2-q^2}\right)\\[0.1cm]
&=\frac{6\,m_N\,m_\pi^2\,|\vec{q}|}{\bigl[(W+m_N)^2-q^2\bigr](m_\pi^2-q^2)}\,F^P(q^2).
\end{split}
\end{equation}

\subsection{Examples}

To observe the behavior of the amplitudes, we have selected two examples: $P_{33} \ (1232)$, the lightest resonance, and $P_{13} \ (1720)$, the heaviest resonance for which no amplitude vanishes, as shown in \Cref{fig:form_factors}. We fix the value of $W$ to the mean value between its minimum and maximum, given an incoming energy of $E_\chi = 10$ GeV in the Lab frame, which results in $W \sim 2.5$ GeV. We do not show the positive spin transitions or all the combinations for the longitudinal ones, as their values are identical to those shown, with only the sign differing. The chosen parameters ensure that the amplitudes do not vanish due to cancellations between the couplings of the up and down quarks: $g_u^k = 2$ and $g_d^k = 1$, where $k = V, A, S, P$ corresponds to the type of coupling. For the dark fermions, the couplings are purely vectorial, so that $g_\chi^k = 1$ for vector couplings and $g_\chi^k = 0$ for axial, scalar, and pseudoscalar couplings.

We observe that a general behavior of the amplitudes  is a decrease of the value as the momentum transfer, $\left| q^2 \right|$, increases. For the vector and axial amplitudes, the transverse polarizations dominate, while the longitudinal exhibits a resonance or pole. This is due to the nature of the MEM, that changes from being  time-like  (lower $\left| q^2 \right|$) to  space-like (higher $\left| q^2 \right|$). The pole occurs  precisely when the MEM is  light-like, leading to higher values for the normalized polarization vector,  $e_s^\mu$. The behavior of the longitudinal polarization at larger $\left| q^2 \right|$ is mainly  governed by the value of the space component of the polarization vector, $e_s^z$, which increases and exhibits a similar shape to that of the amplitudes. 

The present model predicts higher values for the scalar amplitudes, while the pseudoscalar amplitudes are much lower, being suppressed by approximately three orders of magnitude compared to the scalar ones.

\section{Summary}
\label{sec:summary}
This paper developed a comprehensive theoretical framework to describe the single pion production by inelastic interactions between dark fermions and nucleons, in which the scattering process generates resonances that subsequently decay into a nucleon and a pion. Our approach extended the Rein-Sehgal formalism by incorporating multiple mediator types, including dark photons with both vector and axial components, as well as scalar and pseudoscalar interactions in a model-independent way. 

The Rein and Sehgal model is based on the FKR relativistic quark model, which treats hadrons as excitations of a four-dimensional harmonic oscillator.  Detailed derivations of the differential cross sections were presented, regarding the kinematics in the laboratory and isobaric frames. Transition amplitudes for vector, axial, scalar and pseudoscalar interactions were derived by applying appropriate substitutions in the FKR Hamiltonian and incorporating corrections such as the Breit-Winger factor for finite resonance widths and pion-pole contributions dictated by PCAC. The resulting expressions provided a unified description of the resonant dark matter–nucleon interactions at the GeV scale and offered a valuable tool for the analysis and interpretation of experimental data in both dark sector and dark matter searches.

This approach  should be regarded  as an initial  estimation of cross sections, especially suitable for astrophysical scenarios, where precise computations is not as important as in terrestrial experiments. The computation of single pion production by BSM mechanisms will eventually require a more comprehensive modeling  to account for the  complexity of the  hadronic matrix elements. Unfortunately, most recent models rely heavily on experimental data, which lacks for BSM fields. A more detailed approach---fitted carefully for each resonance, along with the inclusion of non-resonant contributions to single pion production---represents a future challenge beyond the scope of the present work.

\section*{acknowledgments}
The authors would like to thank Prof. Krzysztof Graczyk for a useful discussion about the validity of the model considered in the present study. J.~H.~Z. is supported by the National Science Centre, Poland (research grant No. 2021/42/E/ST2/00031). M.~R.~Q. is supported by the Cluster of Excellence “Precision Physics, Fundamental Interactions, and Structure of Matter” (PRISMA+ EXC 2118/1) funded by the Deutsche Forschungsgemeinschaft (DFG, German Research Foundation) within the German Excellence Strategy (Project No. 390831469).

\appendix
\section{List of amplitudes}
\label{sec:table_form_factors}

Here we present the list of amplitudes obtained by computing the transitions from a proton to a resonance mediated by the operators presented above. To obtain those for transitions from a neutron, the up and down couplings simply need to be interchanged.

\vspace{1.5cm}

\LTcapwidth=0.94\textwidth
\begingroup
\renewcommand\arraystretch{2}
\begin{longtable*}[t]{|c|c|c|c|c|c|}
\hline
\centering
\textbf{Resonance} & \boldmath$ \ \ f_{2s} \ \ $ & \textbf{Vector} & \textbf{Axial} & \textbf{Scalar} & \textbf{Pseudoscalar} \ \  \\
\hline
\hline
\multirow{6}{*}{\shortstack{$P_{33} (1232)$\\ $^4(10)_{3/2} [56,0^+]_0$}}
            & $f_{-3}$ & $\sqrt{6} g_1^V R^V$ & $\sqrt{6} g_1^A R^A$ &   &   \\
            & $f_{-1}$ & $\sqrt{2} g_1^V R^V$ & $\sqrt{2} g_1^A R^A$ &   &   \\
            & $f_{+1}$ & $-\sqrt{2} g_1^V R^V$ & $\sqrt{2} g_1^A R^A$ &   &   \\
            & $f_{+3}$ & $-\sqrt{6} g_1^V R^V$ & $\sqrt{6} g_1^A R^A$ &   &   \\
            & $f_{0^+}$ & $0$ & $-2\sqrt{2} g_1^A C$ & $0$ & $\frac{2\sqrt{2}}{9} g_1^P S^P$ \\
            & $f_{0^-}$ & $0$ & $-2\sqrt{2} g_1^A C$ & $0$ & $\frac{2\sqrt{2}}{9} g_1^P S^P$ \\
\hline
\multirow{4}{*}{\shortstack{$P_{11} (1440)$\\ $^2(8)_{1/2} [56,0^+]_2$}}
            & $f_{-1}$ & $\frac{1}{2\sqrt{3}} g_4^V \lambda^2 R^V$ & $\frac{1}{2\sqrt{3}} g_4^A \lambda^2 R^A$ &   &   \\
            & $f_{+1}$ & $\frac{1}{2\sqrt{3}} g_4^V \lambda^2 R^V$ & $-\frac{1}{2\sqrt{3}} g_4^A \lambda^2 R^A$ &   &   \\
            & $f_{0^+}$ & $-\frac{\sqrt{3}}{2} g_5^V \lambda^2 S$ & $\frac{1}{2\sqrt{3}} g_4^A \lambda (\lambda C - 2 B)$ & $-\frac{1}{6\sqrt{3}} g_5^S \lambda^2 S^S$ & $-\frac{1}{18\sqrt{3}} g_4^P \lambda S^P$ \\
            & $f_{0^-}$ & $-\frac{\sqrt{3}}{2} g_5^V \lambda^2 S$ & $-\frac{1}{2\sqrt{3}} g_4^A \lambda (\lambda C - 2 B)$ & $-\frac{1}{6\sqrt{3}} g_5^S \lambda^2 S^S$ & $\frac{1}{18\sqrt{3}} g_4^P \lambda S^P$ \\
\hline
\multirow{9}{*}{\shortstack{$D_{13} (1520)$\\ $^2(8)_{3/2} [70,1^-]_1$}\vspace{2.5cm}}
            & $f_{-3}$ & $-\frac{3}{\sqrt{2}} g_1^V T^V$ & $-\frac{3}{\sqrt{2}} g_1^A T^A$ &   &   \\
            & $f_{-1}$ & $-\frac{3}{\sqrt{2}} g_1^V T^V + \frac{1}{\sqrt{3}} g_2^V \lambda R^V$ & $-\frac{3}{\sqrt{2}} g_1^A T^A + \frac{1}{\sqrt{3}} g_2^A \lambda R^A$ &   &   \\
            & $f_{+1}$ & $-\frac{3}{\sqrt{2}} g_1^V T^V + \frac{1}{\sqrt{3}} g_2^V \lambda R^V$ & $\frac{3}{\sqrt{2}} g_1^A T^A - \frac{1}{\sqrt{3}} g_2^A \lambda R^A$ &   &   \\
            & $f_{+3}$ & $-\frac{3}{\sqrt{2}} g_1^V T^V$ & $\frac{3}{\sqrt{2}} g_1^A T^A$ &   &   \\
            & $f_{0^+}$ & $-\sqrt{3} g_1^V \lambda S$ & $\frac{1}{\sqrt{3}} g_2^A \lambda C$ & $-\frac{1}{3\sqrt{3}} g_1^S \lambda S^S$ & $-\frac{1}{9\sqrt{3}} g_2^P \lambda S^P$ \\
            & $f_{0^-}$ & $-\sqrt{3} g_1^V \lambda S$ & $-\frac{1}{\sqrt{3}} g_2^A \lambda C$ & $-\frac{1}{3\sqrt{3}} g_1^S \lambda S^S$ & $\frac{1}{9\sqrt{3}} g_2^P \lambda S^P$ \\
\hline
\multirow{4}{*}{\shortstack{$S_{11} (1535)$\\ $^2(8)_{1/2} [70,1^-]_1$}}
            & $f_{-1}$ & $-\sqrt{3} g_1^V T^V - \frac{1}{\sqrt{6}} g_2^V \lambda R^V$ & $-\sqrt{3} g_1^A T^A - \frac{1}{\sqrt{6}} g_2^A \lambda R^A$ &   &   \\
            & $f_{+1}$ & $\sqrt{3} g_1^V T^V + \frac{1}{\sqrt{6}} g_2^V \lambda R^V$ & $-\sqrt{3} g_1^A T^A - \frac{1}{\sqrt{6}} g_2^A \lambda R^A$ &   &   \\
            & $f_{0^+}$ & $\sqrt{\frac{3}{2}} g_1^V \lambda S$ & $-\frac{1}{\sqrt{6}} g_2^A (\lambda C - 3 B)$ & $\frac{1}{3\sqrt{6}} g_1^S \lambda S^S$ & $\frac{1}{9\sqrt{6}} g_2^P \lambda S^P$ \\
            & $f_{0^-}$ & $-\sqrt{\frac{3}{2}} g_1^V \lambda S$ & $-\frac{1}{\sqrt{6}} g_2^A (\lambda C - 3 B)$ & $-\frac{1}{3\sqrt{6}} g_1^S \lambda S^S$ & $\frac{1}{9\sqrt{6}} g_2^P \lambda S^P$ \\
\hline
\multirow{6}{*}{\shortstack{$P_{33} (1600)$\\ $^4(10)_{3/2} [56,0^+]_2$}}
            & $f_{-3}$ & $-\frac{1}{\sqrt{2}} g_1^V \lambda^2 R^V$ & $-\frac{1}{\sqrt{2}} g_1^A \lambda^2 R^A$ &   &   \\
            & $f_{-1}$ & $-\frac{1}{\sqrt{6}} g_1^V \lambda^2 R^V$ & $-\frac{1}{\sqrt{6}} g_1^A \lambda^2 R^A$ &   &   \\
            & $f_{+1}$ & $\frac{1}{\sqrt{6}} g_1^V \lambda^2 R^V$ & $-\frac{1}{\sqrt{6}} g_1^A \lambda^2 R^A$ &   &   \\
            & $f_{+3}$ & $\frac{1}{\sqrt{2}} g_1^V \lambda^2 R^V$ & $-\frac{1}{\sqrt{2}} g_1^A \lambda^2 R^A$ &   &   \\
            & $f_{0^+}$ & $0$ & $\sqrt{\frac{2}{3}} g_1^A \lambda (\lambda C - 2 B)$ & $0$ & $-\frac{1}{9} \sqrt{\frac{2}{3}} g_1^P \lambda^2 S^P$ \\
            & $f_{0^-}$ & $0$ & $\sqrt{\frac{2}{3}} g_1^A \lambda (\lambda C - 2 B)$ & $0$ & $-\frac{1}{9} \sqrt{\frac{2}{3}} g_1^P \lambda^2 S^P$ \\
\hline
\multirow{4}{*}{\shortstack{$S_{31} (1620)$\\ $^2(10)_{1/2} [70,1^-]_1$}}
            & $f_{-1}$ & $-\sqrt{3} g_1^V T^V + \frac{1}{\sqrt{6}} g_1^V \lambda R^V$ & $-\sqrt{3} g_1^A T^A + \frac{1}{\sqrt{6}} g_1^A \lambda R^A$ &   &   \\
            & $f_{+1}$ & $\sqrt{3} g_1^V T^V - \frac{1}{\sqrt{6}} g_1^V \lambda R^V$ & $-\sqrt{3} g_1^A T^A + \frac{1}{\sqrt{6}} g_1^A \lambda R^A$ &   &   \\
            & $f_{0^+}$ & $\sqrt{\frac{3}{2}} g_1^V \lambda S$ & $\frac{1}{\sqrt{6}} g_1^A (\lambda C - 3 B)$ & $\frac{1}{3\sqrt{6}} g_1^S \lambda S^S$ & $-\frac{1}{3\sqrt{6}} g_1^P \lambda S^P$ \\
            & $f_{0^-}$ & $-\sqrt{\frac{3}{2}} g_1^V \lambda S$ & $\frac{1}{\sqrt{6}} g_1^A (\lambda C - 3 B)$ & $-\frac{1}{3\sqrt{6}} g_1^S \lambda S^S$ & $-\frac{1}{3\sqrt{6}} g_1^P \lambda S^P$ \\
\hline
\multirow{4}{*}{\shortstack{$S_{11} (1650)$\\ $^4(8)_{1/2} [70,1^-]_1$}}
            & $f_{-1}$ & $\frac{1}{\sqrt{6}} g_3^V \lambda R^V$ & $\frac{1}{\sqrt{6}} g_3^A \lambda R^A$ &   &   \\
            & $f_{+1}$ & $-\frac{1}{\sqrt{6}} g_3^V \lambda R^V$ & $\frac{1}{\sqrt{6}} g_3^A \lambda R^A$ &   &   \\
            & $f_{0^+}$ & $0$ & $-\sqrt{\frac{2}{3}} g_3^A (\lambda C - 3 B)$ & $0$ & $\frac{1}{9} \sqrt{\frac{2}{3}} g_3^P \lambda S^P$ \\
            & $f_{0^-}$ & $0$ & $-\sqrt{\frac{2}{3}} g_3^A (\lambda C - 3 B)$ & $0$ & $\frac{1}{9} \sqrt{\frac{2}{3}} g_3^P \lambda S^P$ \\
\hline
\multirow{4}{*}{\shortstack{$D_{15} (1675)$\\ $^4(8)_{5/2} [70,1^-]_1$}}
            & $f_{-3}$ & $-\sqrt{\frac{3}{5}} g_3^V \lambda R^V$ & $-\sqrt{\frac{3}{5}} g_3^A \lambda R^A$ &   &   \\
            & $f_{-1}$ & $-\sqrt{\frac{3}{10}} g_3^V \lambda R^V$ & $-\sqrt{\frac{3}{10}} g_3^A \lambda R^A$ &   &   \\
            & $f_{+1}$ & $\sqrt{\frac{3}{10}} g_3^V \lambda R^V$ & $-\sqrt{\frac{3}{10}} g_3^A \lambda R^A$ &   &   \\
            & $f_{+3}$ & $\sqrt{\frac{3}{5}} g_3^V \lambda R^V$ & $-\sqrt{\frac{3}{5}} g_3^A \lambda R^A$ &   &   \\
            & $f_{0^+}$ & $0$ & $\sqrt{\frac{6}{5}} g_3^A \lambda C$ & $0$ & $-\frac{1}{3} \sqrt{\frac{2}{15}} g_3^P \lambda S^P$ \\
            & $f_{0^-}$ & $0$ & $\sqrt{\frac{6}{5}} g_3^A \lambda C$ & $0$ & $-\frac{1}{3} \sqrt{\frac{2}{15}} g_3^P \lambda S^P$ \\
\hline
\multirow{6}{*}{\shortstack{$F_{15} (1680)$\\ $^2(8)_{5/2} [56,2^+]_2$}}
            & $f_{-3}$ & $3\sqrt{\frac{2}{5}} g_5^V \lambda T^V$ & $3\sqrt{\frac{2}{5}} g_5^A \lambda T^A$ &   &   \\
            & $f_{-1}$ & $\frac{3}{\sqrt{5}} g_5^V \lambda T^V - \frac{1}{\sqrt{10}} g_4^V \lambda^2 R^V$ & $\frac{3}{\sqrt{5}} g_5^A \lambda T^A - \frac{1}{\sqrt{10}} g_4^A \lambda^2 R^A$ &   &   \\
            & $f_{+1}$ & $\frac{3}{\sqrt{5}} g_5^V \lambda T^V - \frac{1}{\sqrt{10}} g_4^V \lambda^2 R^V$ & $-\frac{3}{\sqrt{5}} g_5^A \lambda T^A + \frac{1}{\sqrt{10}} g_4^A \lambda^2 R^A$ &   &   \\
            & $f_{+3}$ & $3\sqrt{\frac{2}{5}} g_5^V \lambda T^V$ & $-3\sqrt{\frac{2}{5}} g_5^A \lambda T^A$ &   &   \\
            & $f_{0^+}$ & $\frac{3}{\sqrt{10}} g_5^V \lambda^2 S$ & $-\sqrt{\frac{2}{5}}  g_4^A \lambda^2 C$ & $\frac{1}{3\sqrt{10}} g_5^S \lambda^2 S^S$ & $\frac{1}{9\sqrt{10}} g_4^P \lambda^2 S^P$ \\
            & $f_{0^-}$ & $\frac{3}{\sqrt{10}} g_5^V \lambda^2 S$ & $\sqrt{\frac{2}{5}} g_4^A \lambda^2 C$ & $\frac{1}{3\sqrt{10}} g_5^S \lambda^2 S^S$ & $-\frac{1}{9\sqrt{10}} g_4^P \lambda^2 S^P$ \\
\hline
\multirow{6}{*}{\shortstack{$D_{13} (1700)$\\ $^4(8)_{3/2} [70,1^-]_1$}}
            & $f_{-3}$ & $\frac{3}{\sqrt{10}} g_3^V \lambda R^V$ & $\frac{3}{\sqrt{10}} g_3^A \lambda R^A$ &   &   \\
            & $f_{-1}$ & $\frac{1}{\sqrt{30}} g_3^V \lambda R^V$ & $\frac{1}{\sqrt{30}} g_3^A \lambda R^A$ &   &   \\
            & $f_{+1}$ & $\frac{1}{\sqrt{30}} g_3^V \lambda R^V$ & $-\frac{1}{\sqrt{30}} g_3^A \lambda R^A$ &   &   \\
            & $f_{+3}$ & $\frac{3}{\sqrt{10}} g_3^V \lambda R^V$ & $-\frac{3}{\sqrt{10}} g_3^A \lambda R^A$ &   &   \\
            & $f_{0^+}$ & $0$ & $-\sqrt{\frac{2}{15}} g_3^A \lambda C$ & $0$ & $\frac{1}{9} \sqrt{\frac{2}{15}} g_3^P \lambda S^P$ \\
            & $f_{0^-}$ & $0$ & $\sqrt{\frac{2}{15}} g_3^A \lambda C$ & $0$ & $-\frac{1}{9} \sqrt{\frac{2}{15}} g_3^P \lambda S^P$ \\
\hline
\multirow{6}{*}{\shortstack{$D_{33} (1700)$\\ $^2(10)_{3/2} [70,1^-]_1$}}
            & $f_{-3}$ & $-\frac{3}{\sqrt{2}} g_1^V T^V$ & $-\frac{3}{\sqrt{2}} g_1^A T^A$ &   &   \\
            & $f_{-1}$ & $-\sqrt{\frac{3}{2}} g_1^V T^V - \frac{1}{\sqrt{3}} g_1^V \lambda R^V$ & $-\sqrt{\frac{3}{2}} g_1^A T^A - \frac{1}{\sqrt{3}} g_1^A \lambda R^A$ &   &   \\
            & $f_{+1}$ & $-\sqrt{\frac{3}{2}} g_1^V T^V - \frac{1}{\sqrt{3}} g_1^V \lambda R^V$ & $\sqrt{\frac{3}{2}} g_1^A T^A + \frac{1}{\sqrt{3}} g_1^A \lambda R^A$ &   &   \\
            & $f_{+3}$ & $-\frac{3}{\sqrt{2}} g_1^V T^V$ & $\frac{3}{\sqrt{2}} g_1^A T^A$ &   &   \\
            & $f_{0^+}$ & $-\sqrt{3} g_1^V \lambda S$ & $-\frac{1}{\sqrt{3}} g_1^A \lambda C$ & $-\frac{1}{3\sqrt{3}} g_1^S \lambda S^S$ & $\frac{1}{9\sqrt{3}} g_1^P \lambda S^P$ \\
            & $f_{0^-}$ & $-\sqrt{3} g_1^V \lambda S$ & $\frac{1}{\sqrt{3}} g_1^A \lambda C$ & $-\frac{1}{3\sqrt{3}} g_1^S \lambda S^S$ & $-\frac{1}{9\sqrt{3}} g_1^P \lambda S^P$ \\
\hline
\multirow{4}{*}{\shortstack{$P_{11} (1710)$\\ $^2(8)_{1/2} [70,0^+]_2$}}
            & $f_{-1}$ & $-\frac{1}{2\sqrt{6}} g_2^V \lambda^2 R^V$ & $-\frac{1}{2\sqrt{6}} g_2^A \lambda^2 R^A$ &   &   \\
            & $f_{+1}$ & $-\frac{1}{2\sqrt{6}} g_2^V \lambda^2 R^V$ & $\frac{1}{2\sqrt{6}} g_2^A \lambda^2 R^A$ &   &   \\
            & $f_{0^+}$ & $\frac{1}{2} \sqrt{\frac{3}{2}} g_1^V \lambda^2 S$ & $-\frac{1}{2\sqrt{6}} g_2^A \lambda(\lambda C - 2B)$ & $\frac{1}{6\sqrt{6}} g_1^S \lambda^2 S^S$ & $\frac{1}{18\sqrt{6}} g_2^P \lambda^2 S^P$ \\
            & $f_{0^-}$ & $\frac{1}{2} \sqrt{\frac{3}{2}} g_1^V \lambda^2 S$ & $\frac{1}{2\sqrt{6}} g_2^A \lambda(\lambda C - 2B)$ & $\frac{1}{6\sqrt{6}} g_1^S \lambda^2 S^S$ & $-\frac{1}{18\sqrt{6}} g_2^P \lambda^2 S^P$ \\
\hline
\multirow{6}{*}{\shortstack{$P_{13} (1720)$\\ $^2(8)_{3/2} [56,2^+]_2$}}
            & $f_{-3}$ & $-\frac{3}{\sqrt{10}} g_5^V \lambda T^V$ & $-\frac{3}{\sqrt{10}} g_5^A \lambda T^A$ &   &   \\
            & $f_{-1}$ & $3\sqrt{\frac{3}{10}} g_5^V \lambda T^V + \frac{1}{\sqrt{15}} g_4^V \lambda^2 R^V$ & $3\sqrt{\frac{3}{10}} g_5^A \lambda T^A + \frac{1}{\sqrt{15}} g_4^A \lambda^2 R^A$ &   &   \\
            & $f_{+1}$ & $-3\sqrt{\frac{3}{10}} g_5^V \lambda T^V - \frac{1}{\sqrt{15}} g_4^V \lambda^2 R^V$ & $3\sqrt{\frac{3}{10}} g_5^A \lambda T^A + \frac{1}{\sqrt{15}} g_4^A \lambda^2 R^A$ &   &   \\
            & $f_{+3}$ & $\frac{3}{\sqrt{10}} g_5^V \lambda T^V$ & $-\frac{3}{\sqrt{10}} g_5^A \lambda T^A$ &   &   \\
            & $f_{0^+}$ & $-\sqrt{\frac{3}{5}} g_5^V \lambda^2 S$ & $\frac{1}{\sqrt{15}} g_4^A \lambda (\lambda C - 5 B)$ & $-\frac{1}{3\sqrt{15}} g_5^S \lambda^2 S^S$ & $-\frac{1}{9\sqrt{15}} g_4^P \lambda^2 S^P$ \\
            & $f_{0^-}$ & $\sqrt{\frac{3}{5}} g_5^V \lambda^2 S$ & $\frac{1}{\sqrt{15}} g_4^A \lambda (\lambda C - 5 B)$ & $\frac{1}{3\sqrt{15}} g_5^S \lambda^2 S^S$ & $-\frac{1}{9\sqrt{15}} g_4^P \lambda^2 S^P$ \\
            & & & & &\\
\hline
\multirow{6}{*}{\shortstack{$F_{35} (1905)$\\ $^4(10)_{5/2} [56,2^+]_2$}}
            & $f_{-3}$ & $-3\sqrt{\frac{2}{35}} g_1^V \lambda^2 R^V$ & $-3\sqrt{\frac{2}{35}} g_1^A \lambda^2 R^A$ &   &   \\
            & $f_{-1}$ & $-\sqrt{\frac{1}{35}} g_1^V \lambda^2 R^V$ & $-\sqrt{\frac{1}{35}} g_1^A \lambda^2 R^A$ &   &   \\
            & $f_{+1}$ & $-\sqrt{\frac{1}{35}} g_1^V \lambda^2 R^V$ & $\sqrt{\frac{1}{35}} g_1^A \lambda^2 R^A$ &   &   \\
            & $f_{+3}$ & $-3\sqrt{\frac{2}{35}} g_1^V \lambda^2 R^V$ & $3\sqrt{\frac{2}{35}} g_1^A \lambda^2 R^A$ &   &   \\
            & $f_{0^+}$ & $0$ & $\frac{2}{\sqrt{35}} g_1^A \lambda^2 C$ & $0$ & $-\frac{2}{9\sqrt{35}} g_1^P \lambda^2 S^P$ \\
            & $f_{0^-}$ & $0$ & $-\frac{2}{\sqrt{35}} g_1^A \lambda^2 C$ & $0$ & $\frac{2}{9\sqrt{35}} g_1^P \lambda^2 S^P$ \\
\hline
\multirow{4}{*}{\shortstack{$P_{31} (1910)$\\ $^4(10)_{1/2} [56,2^+]_2$}\vspace{1.5cm}}
            & $f_{-1}$ & $\frac{1}{\sqrt{15}} g_1^V \lambda^2 R^V$ & $\frac{1}{\sqrt{15}} g_1^A \lambda^2 R^A$ &   &   \\
            & $f_{+1}$ & $\frac{1}{\sqrt{15}} g_1^V \lambda^2 R^V$ & $-\frac{1}{\sqrt{15}} g_1^A \lambda^2 R^A$ &   &   \\
            & $f_{0^+}$ & $0$ & $-\frac{2}{\sqrt{15}} g_1^A \lambda ( \lambda C - 5 B)$ & $0$ & $\frac{2}{9\sqrt{15}} g_1^P \lambda^2 S^P$ \\
            & $f_{0^-}$ & $0$ & $\frac{2}{\sqrt{15}} g_1^A \lambda ( \lambda C - 5 B)$ & $0$ & $-\frac{2}{9\sqrt{15}} g_1^P \lambda^2 S^P$ \\
\hline
\multirow{6}{*}{\shortstack{$P_{33} (1920)$\\ $^4(10)_{3/2} [56,2^+]_2$}}
            & $f_{-3}$ & $\frac{1}{\sqrt{5}} g_1^V \lambda^2 R^V$ & $\frac{1}{\sqrt{5}} g_1^A \lambda^2 R^A$ &   &   \\
            & $f_{-1}$ & $-\frac{1}{\sqrt{15}} g_1^V \lambda^2 R^V$ & $-\frac{1}{\sqrt{15}} g_1^A \lambda^2 R^A$ &   &   \\
            & $f_{+1}$ & $\frac{1}{\sqrt{15}} g_1^V \lambda^2 R^V$ & $-\frac{1}{\sqrt{15}} g_1^A \lambda^2 R^A$ &   &   \\
            & $f_{+3}$ & $-\frac{1}{\sqrt{5}} g_1^V \lambda^2 R^V$ & $\frac{1}{\sqrt{5}} g_1^A \lambda^2 R^A$ &   &   \\
            & $f_{0^+}$ & $0$ & $\frac{2}{\sqrt{15}} g_1^A \lambda ( \lambda C - 5 B)$ & $0$ & $-\frac{2}{9\sqrt{15}} g_1^P \lambda^2 S^P$ \\
            & $f_{0^-}$ & $0$ & $\frac{2}{\sqrt{15}} g_1^A \lambda ( \lambda C - 5 B)$ & $0$ & $-\frac{2}{9\sqrt{15}} g_1^P \lambda^2 S^P$ \\
\hline
\multirow{6}{*}{\shortstack{$F_{37} (1950)$\\ $^4(10)_{7/2} [56,2^+]_2$}}
            & $f_{-3}$ & $\sqrt{\frac{2}{7}} g_1^V \lambda^2 R^V$ & $\sqrt{\frac{2}{7}} g_1^A \lambda^2 R^A$ &   &   \\
            & $f_{-1}$ & $\sqrt{\frac{6}{35}} g_1^V \lambda^2 R^V$ & $\sqrt{\frac{6}{35}} g_1^A \lambda^2 R^A$ &   &   \\
            & $f_{+1}$ & $-\sqrt{\frac{6}{35}} g_1^V \lambda^2 R^V$ & $\sqrt{\frac{6}{35}} g_1^A \lambda^2 R^A$ &   &   \\
            & $f_{+3}$ & $-\sqrt{\frac{2}{7}} g_1^V \lambda^2 R^V$ & $\sqrt{\frac{2}{7}} g_1^A \lambda^2 R^A$ &   &   \\
            & $f_{0^+}$ & $0$ & $-2\sqrt{\frac{6}{35}} g_1^A \lambda^2 C$ & $0$ & $\frac{2}{3} \sqrt{\frac{2}{105}} g_1^P \lambda^2 S^P$ \\
            & $f_{0^-}$ & $0$ & $-2\sqrt{\frac{6}{35}} g_1^A \lambda^2 C$ & $0$ & $\frac{2}{3} \sqrt{\frac{2}{105}} g_1^P \lambda^2 S^P$ \\
\hline

\multirow{6}{*}{\shortstack{$F_{17} (1970)$\\ $^4(8)_{7/2} [70,2^+]_2$}}
            & $f_{-3}$ & $-\frac{1}{\sqrt{7}} g_3^V \lambda^2 R^V$ & $-\frac{1}{\sqrt{7}} g_3^A \lambda^2 R^A$ &   &   \\
            & $f_{-1}$ & $-\sqrt{\frac{3}{35}} g_3^V \lambda^2 R^V$ & $-\sqrt{\frac{3}{35}} g_3^A \lambda^2 R^A$ &   &   \\
            & $f_{+1}$ & $\sqrt{\frac{3}{35}} g_3^V \lambda^2 R^V$ & $-\sqrt{\frac{3}{35}} g_3^A \lambda^2 R^A$ &   &   \\
            & $f_{+3}$ & $\frac{1}{\sqrt{7}} g_3^V \lambda^2 R^V$ & $-\frac{1}{\sqrt{7}} g_3^A \lambda^2 R^A$ &   &   \\
            & $f_{0^+}$ & $0$ & $2\sqrt{\frac{3}{35}} g_3^A \lambda^2 C$ & $0$ & $-\frac{2}{3\sqrt{105}} g_3^P \lambda^2 S^P$ \\
            & $f_{0^-}$ & $0$ & $2\sqrt{\frac{3}{35}} g_3^A \lambda^2 C$ & $0$ & $-\frac{2}{3\sqrt{105}} g_3^P \lambda^2 S^P$ \\
\hline
\caption{Amplitudes for transitions from protons to resonances. We use some short notation for the couplings, such that for $k \equiv V,A,S,P$: $g^k_1 \equiv \left(g_u^k - g_d^k \right)$, $g^k_2 \equiv \left(5g_u^k + g_d^k \right)$, $g^k_3 \equiv \left(g_u^k + 2g_d^k \right)$, $g^k_4 \equiv \left(4g_u^k - g_d^k \right)$, $g^k_5 \equiv \left(2g_u^k + g_d^k \right)$. For neutrons, just interchange the couplings $g_u^k \leftrightarrow g_d^k$ in the previous definitions of $g^k_i$. It should also be noticed that we are not putting the explicit dependence on the dark sector helicity combinations, $\lambda_1, \lambda_2$.}
\label{tab:form_factors}

\end{longtable*}
\endgroup

\section{Most representative quantities}
\label{sec:important_quantities}

\response{ Here we briefly present a condensed summary of the main quantities computed and used throughout the text to facilitate the reading of the current work.}

\subsection{Kinematics}

\response{ We call $p_1$ ($p_3$) the 4-momentum of the incoming (outgoing) $\chi$, while $p_2$ ($p_4$) the one of the nucleon target (resonance produced), all in the Lab frame. $q$ is the transferred 4-momentum, from the DM to the hadronic current. These quantities are parametrized as follows:}
\begin{equation}
\begin{split}
&p_1 = \left( E_1, 0, 0, |\vec{p}_1|\right),\\[0.05cm]
&p_2 = \left( m_N, 0, 0, 0\right),\\[0.05cm]
&p_3 = \left( E_3, |\vec{p}_3| \sin \theta, 0, |\vec{p}_3| \cos \theta\right),\\[0.05cm]
&p_4 = p_1 + p_2 - p_3,\\[0.05cm]
&q = p_1 - p_2\,.
\end{split}
\end{equation}
\response{ In the IB frame, those kinematical quantities can be found, parametrized by the kinematical variables used in the Lab frame:}
\begin{widetext}
\begin{equation}
\begin{split}
&p_1^\mathrm{IB} = \left(\frac{(E_1 - E_3)^2 + 2 m_N E_1 - Q^2}{2W} , A_{13}, 0, B_{13}^-\right),\\[0.1cm]
&p_2^\mathrm{IB} = \left(\frac{m_N(E_1 - E_3 + m_N)}{W}, 0, 0, -\frac{ m_N Q }{W}\right),\\[0.1cm]
&p_3^\mathrm{IB} = \left(-\frac{(E_1 - E_3)^2 - 2 m_N E_3 - Q^2}{2W}, A_{13}, 0, B_{13}^+\right),\\[0.1cm]
&p_4^\mathrm{IB} = \left(W, 0, 0, 0\right),\\[0.1cm]
&A_{13} =  \frac{\sqrt{\left(|\vec{p}_1| + |\vec{p}_3| - Q\right)\left(|\vec{p}_1| - |\vec{p}_3| + Q\right)\left(-|\vec{p}_1| + |\vec{p}_3| + Q\right)\left(|\vec{p}_1| + |\vec{p}_3| + Q\right))} }{2Q}, \\[0.1cm]
&B_{13}^{\pm} = \frac{(E_1^2 - E_3^2)(E_1 - E_3 + m_N) - (E_1 + E_3 \pm m_N) Q^2}{2QW}\,,
\end{split}
\end{equation}
\end{widetext}

\response{For vector - axial couplings, the unit vectors used to parametrize each helicity component of the amplitude transitions are:}
\begin{equation}
\begin{split}
e_L^\mu &= \frac{1}{\sqrt{2}} \left( 0, 1, -i, 0 \right),\\
e_R^\mu &= \frac{1}{\sqrt{2}} \left( 0, -1, -i, 0 \right),\\
e_{s}^\mu(\lambda_1 \lambda_2) &= \frac{( V_{\lambda_1 \lambda_2}^0, 0, 0, V_{\lambda_1 \lambda_2}^3 )}{\sqrt{\left| \left(V_{\lambda_1 \lambda_2}^0 \right)^2 - \left(V_{\lambda_1 \lambda_2}^3 \right)^2 \right|}}\,,
\end{split}
\end{equation}
\response{where $V_{\lambda_1 \lambda_2}$ is the MEM. }

\subsection{Helicity amplitudes}

\response{Now, we present the parameters needed to compute the helicity amplitudes , as worked out in \Cref{sec:form-factors}. A parameter used for any kind of interaction is $\lambda$, defined as:}
\begin{equation}
\begin{split}
\lambda &\equiv \sqrt{\frac{2}{\Omega}} \frac{m_N}{W} |\vec{q}|\,.
\end{split}
\end{equation}

\response{For the vectorial case, we have:}
\begin{equation}
\begin{split}
\lambda &\equiv \sqrt{\frac{2}{\Omega}} \frac{m_N}{W} |\vec{q}|,\\
R^V &\equiv \frac{\sqrt{2} m_N}{W} \frac{\left(W + m_N \right) |\vec{q}|}{\left(W + m_N \right)^2 - q^2} F^V (q^2),\\
T^V &\equiv \frac{\sqrt{\Omega} }{3 \sqrt{2} W} F^V (q^2),\\
S_{\lambda_1 \lambda_2} &\equiv \frac{m_N}{W \left|\vec{q}\right|} \frac{V_{\lambda_1 \lambda_2}^3 q^0_\mathrm{IB} - V_{\lambda_1 \lambda_2}^0 \left|\vec{q}_\mathrm{IB}\right|}{C_s^{\lambda_1 \lambda_2}} \\
& \ \ \ \times \left(\frac{1}{6} - \frac{q^2}{6 m_N^2} - \frac{W}{2m_N} \right) F^V (q^2)\,.
\end{split}
\end{equation}

\response{For the axial case, we have:}
\begin{align*}
R^A \!\equiv& \frac{\sqrt{2}Z}{6 W}\! \left(W \!+\! m_N \!+ \!\frac{2 N \Omega W}{(W + m_N)^2 - q^2 } \right)\!  F^A (q^2),\\[0.1cm]
T^A \!\equiv &\frac{\sqrt{2\Omega} Z}{3} \frac{m_N |\vec{q}|}{W \left((W + m_N)^2 - q^2 \right)} F^A (q^2),\\[0.1cm]
B_{\lambda_1 \lambda_2} \!\equiv &\frac{1}{C_s^{\lambda_1 \lambda_2}} \sqrt{\frac{\Omega}{2}}\! \left(V_{\lambda_1 \lambda_2}^0 \!\!+\! V_{\lambda_1 \lambda_2}^3 \frac{|\vec{q}^{\,\mathrm{IB}}|}{a m_N} \right)\! Z \frac{F^A (q^2)}{3 W},\nonumber
\end{align*}
\begin{equation}
\begin{split}
C_{\lambda_1 \lambda_2} \!\equiv& \frac{1}{C_s^{\lambda_1 \lambda_2}} \!\bigg[ ( V_{\lambda_1 \lambda_2}^0 |\vec{q}^{\,\mathrm{IB}}|\! -\! V_{\lambda_1 \lambda_2}^3 q_0^\mathrm{IB} )\!\! \left(\frac{1}{3} \!+\! \frac{q_0^\mathrm{IB}}{a m_N} \right) \\[0.1cm]
  +& V_{\lambda_1 \lambda_2}^3 \left( \frac{2}{3} W + \frac{q^2}{a m_N} + \frac{N \Omega}{3 a m_N} \right) \bigg] \frac{Z}{2 W} F^A (q^2),\\[0.1cm]
 a \equiv &1 + \frac{W^2 + m_N^2 - q^2}{2 m_N W}.
\end{split}
\end{equation}
\response{where the last two parameters need to be redefined in order to account for the pion-pole term that appears due to the partial conservation of the axial current:}
\begin{equation}
\begin{split}
B_{\lambda_1 \lambda_2}\!\!\! \to &B_{\lambda_1 \lambda_2}\! +\! \frac{1}{C_s^{\lambda_1 \lambda_2}} \frac{Z}{2 W } \left( V_{\lambda_1 \lambda_2}^0 q_0^\mathrm{IB} \!-\! V_{\lambda_1 \lambda_2}^3 |\vec{q}^{\,\mathrm{IB}}| \right) \\[0.1cm]
&\times \frac{\frac{2}{3}\sqrt{\frac{\Omega}{2}} \left( q_0^\mathrm{IB} + \frac{|\vec{q}^{\,\mathrm{IB}}|^2}{a m_N} \right)}{m_\pi^2 - q^2} F^A (q^2),\\[0.1cm]
 C_{\lambda_1 \lambda_2}\!\!\! \to &C_{\lambda_1 \lambda_2} \!+\!  \frac{1}{C_s^{\lambda_1 \lambda_2}} \frac{Z}{2 W}  \left( V_{\lambda_1 \lambda_2}^0 q_0^\mathrm{IB} \!-\! V_{\lambda_1 \lambda_2}^3 |\vec{q}^{\,\mathrm{IB}}| \right) \\[0.1cm]
 &\times \frac{|\vec{q}^{\,\mathrm{IB}}| \left( \frac{2}{3} W + \frac{q^2}{a m_N} + \frac{N \Omega}{3 a m_N} \right)}{m_\pi^2 - q^2}  F^A (q^2)\,.
\end{split}
\end{equation}

\response{The scalar and the pseudoscalar cases exhibit a much simpler parameter structure, due to the also simpler Lorentz structure of their interactions and because of just being able to mediate states with equal spins. In these cases,  we have:}
\begin{equation}
\begin{split}
S^S &\equiv 3\,F^S(q^2)\,\\
S^P &\equiv \frac{6\,m_N\,|\vec{q}|}{(W+m_N)^2-q^2}\,F^P(q^2)\,.
\end{split}
\end{equation}

\response{With all of these quantities we can obtain the helicity amplitudes found in \Cref{tab:form_factors}.
Finally, we also need the form factors, that take the following forms:}
\begin{equation}
\begin{split}
F^V (q^2) = &\frac{1}{2} \sqrt{1 - \frac{q^2}{(m_N+W)^2}} \left(1 - \frac{q^2}{4 m_N^2} \right)^{-N} \\[0.1 cm]& \times \sqrt{3 \left( G_3 (q^2) \right)^2 + \left( G_1 (q^2) \right)^2}\,,\\
F^A (q^2) = &\frac{\sqrt{3}}{2} \sqrt{1 - \frac{q^2}{(m_N+W)^2}} \left(1 - \frac{q^2}{4 m_N^2} \right)^{-N} \\[0.1cm]
& \times \left[1 - \frac{W^2 + q^2 - m_N^2}{8 m_N^2} \right] C_5^A (q^2)\,\\
F^S(q^2) = &\frac{F^S(0)}{1 - \frac{q^2}{m_S^2}}\,,\\
F^P(q^2) = &\frac{F^P(0)}{1-\frac{q^2}{m_P^2}}\,.
\end{split}
\end{equation}

\response{The definitions of the quantities therein involved can be found in \Cref{sec:form-factors}.}
\vspace{3cm}

\bibliographystyle{apsrev4-1}
\bibliography{lib}

\end{document}